\documentclass{aastex}
\begin{document}
\singlespace
\title{Sloshing in High Speed Galaxy Interactions}
\shorttitle{Sloshing in High Speed Galaxy Interactions}
\shortauthors{Zeltwanger et al.}

\author{Thomas Zeltwanger}
\affil{University of Maine}
\affil{Department of Physics and Astronomy, 5709 Bennett Hall,
Orono, ME 04469}
\email{kickers@email.com}
\author{Neil F. Comins}
\affil{University of Maine}
\affil{Department of Physics and Astronomy, 5709 Bennett Hall,
Orono, ME 04469}
\email{neil.comins@umit.maine.edu}
\and
\author{Richard V. E. Lovelace}
\affil {Cornell University, Department of Astronomy,
Ithaca, NY, 14853}
\email{rvl1@Cornell.edu}

\begin{abstract}
Observations of lopsided spiral galaxies motivated us to explore
whether the rapid passage of a companion galaxy could cause them.
We
examine whether the center of mass of the visible matter becomes
displaced from the center of mass of the dark halo during the
intruder's passage, thereby causing an asymmetric response and
asymmetric structure. Two dimensional $N$-body simulations
indicate that this can happen.  
     
We also explore some consequences of this offset. These include
the
center of mass of the visible disk following a decaying orbit
around the halo center of mass and the development of transient
one-armed spirals that persist for up to six rotation periods.

We then study the results of a variety of initial conditions
based on
such offsets.  We report on the results of several runs in which
we
initially offset a disk from its halo's center of mass by an
amount
typical of the above interaction.  In some runs the halo is free
to move, while in others it is held fixed.  We used three
different mass
distributions for the halo in these runs.  We find that the
disk's
center of mass spiraled inward creating a variety of observed or
observable phenomena including one-armed spirals, massive clumps
of
particles, and counter-rotating waves.  The systems settle into
relatively axisymmetric configurations.  Whether or not the end
states
included a bar depended on a variety of initial conditions.
\end{abstract}

\keywords{galaxies: interactions --- galaxies: spiral ---
methods: n-body simulations}

\section{INTRODUCTION}

 Several groups have reported observations of lopsided \ion{H}{1}
\citep{Baldwinetal1980, RichterSancisi1994} and stellar
distributions
\citep{Laueretal1993, Laueretal1996,
RixZaritsky1995, Davidgeetal1997, ZaritskyRix1997, Conselice1997}
in a significant fraction of spiral galaxies.  The
percentage of galaxies that are measurably lopsided range from
30\% of field spirals  \citep{ZaritskyRix1997, Kornreichetal1998} to 50\% of all
spiral galaxies  \citep{Swatersetal1999}. These numbers may be
high if,
as proposed by \citet{ZaritskyRix1997}, lopsided galaxies are
brighter
than symmetric galaxies, thereby biasing the statistics.
Nevertheless,
a significant number of
lopsided galaxies are observed.  These findings have prompted a
number of papers exploring the origins and dynamics of these
asymmetries.

\citet{Baldwinetal1980} proposed a simple kinematic model for
lopsided
galaxies in which different rings of the galaxy, assumed
non-interacting, are initially shifted from their centered
equilibrium positions.
   The shifted rings precess in the overall gravitational
potential in
a direction opposite to that of the mass motion.
   Because the precession rate decreases in general with radial
distance, an initial disturbance tends to  ``wind up" into a
leading
spiral arm in a time appreciably less than the Hubble time.
\citet{MillerSmith1988, MillerSmith1992} have made extensive
computer
simulation studies of the unstable eccentric motion of matter in
the
nuclei of galaxies, which they suggest is pertinent to the
off-center
nuclei observed in a number of galaxies such as M31, M33, and
M101.

More recent ideas on the origin of lopsidedness include galaxy
mergers
\citep{ZaritskyRix1997}, and counter-rotating matter
\citep{SellwoodMerritt1994}. \citet{Schoenmakersetal1997}
interpret optical asymmetry as an indicator of asymmetry in the
overall
galactic potential and therefore an indicator of the  spatial
distribution of the dark matter in a galaxy, which may have a
triaxial
distribution. By analyzing the spiral components present in the
surface
brightness or \ion{H}{1} distribution, the velocity gradient, and
therefore, the shape of the gravitational potential, may be
uncovered. 
\citet{Jog1997} studied  the orbits of stars and gas in lopsided
potentials and concluded that a stationary lopsided disk may
indicate
asymmetry in the halo.

    A further possibility is that an optical disk may be in a
quasi-stationary lopsided state in a symmetric potential, as
discussed
by \citet{SyerTremaine1996}.
   In this model, gaseous and stellar matter that is not fully
relaxed swirl about the
minimum of the halo potential.
   The result is a lopsided flow within a symmetric mass
distribution.
    Numerical simulations of this situation have been done by
\citet{LevineSparke1998} using a gravitational $N$-body tree-code
method for  disk galaxies shifted from  the center of the main
halo
potential. The results are suggestive of lopsidedness with large
lifetimes.

   An $N$-body simulation study of a rotating spheroidal stellar
system
including the dynamics of a massive central object by
\citet{TagaIye1998a} indicates that the central object goes into
a long
lasting oscillation similar to those found earlier by
\citet{MillerSmith1988, MillerSmith1992} and which may explain
asymmetric structures observed in M31 and NGC 4486B\@.
   A linear stability analysis of a self-gravitating fluid disk
with a massive central object also by \citet{TagaIye1998b}
indicates a linear instability \citep{TagaIye1998b}. 
     A study of the eccentric motion of a disk galaxy made up of
a
large number of self-gravitating rings shows instability in the
inner
part of the disk ($<2$ kpc) and the excitation of long lasting
trailing
one armed spiral waves \citep{Lovelaceetal1999}.

In this paper we explore two issues surrounding lopsided
galaxies.  First we examine the gravitational response of an
$N$-body disk
galaxy model embedded in a halo to the high speed passage of a
companion galaxy.  Then, assuming that a disk galaxy has been
perturbed
by a passing galaxy or other mechanism, we explore the response
of the system to a variety of
displacements and relative velocities of a disk of stars offset from
their
halo centers of mass.  In some of these runs, we have both the
stars
and halo free to move, while the halo is fixed in the remaining
runs.  We report on runs
with different initial conditions and different mass
distributions.

We use our two dimensional $N$-body GALAXY code
\citep{SchroederComins1989, Schroeder1989, Shorey1996}, with
100,000
collisionless particles representing stars (or perhaps more
properly, galactic clusters) and a gravitating halo that
comprises
75\% of the system's total mass.  One of our goals is to determine how
perturbations, such as the passage of a companion galaxy or offset halo-disk
centers of mass, affect the formation and evolution of bars in galaxies.  A
sufficiently massive halo will quickly dampen almost any bar
\citep{BinneyTremaine1987,Sellwood1983}.
We have therefore chosen this $75\%$ halo because it has insufficient mass to
suppress the development of a bar that may be driven into existence during the
run \citep{Cominsetal1997}.  The halos we use are sufficiently massive,
however, to simulate the
gravitational effect of a significant dark matter galactic component.  We
therefore expect that the suppression of any bar that develops and decays in
our runs would be due to the dynamics that the disk undergoes.

Both the creation and the destruction of bars have been discussed by
several groups. \citet{Noguchi1996} proposed that bars in late-type galaxies
have intrinsic origins, while bars in early-type galaxies have been formed
by tidal interactions with other galaxies.  Specifically,
\citet{MiwaNoguchi1998} found that tidal effects due to a close encounter
between galaxies may trigger the formation of a bar in an otherwise
barstable disk.
\citet{Athanassoula1996a,Athanassoula1996b,Athanassoula1999} performed
$N$-body simulations of the interactions of a barred disk galaxy with a close
companion galaxy.  She found that the bar may be destroyed as a consequence
of this interaction, or that an offcentered bar may form.  The destruction of
the bar occurs most likely in the case of a merger, when most of the companion
reaches the target's center.

The initial particle velocity distribution in each run is determined by
$Q_{0}$,
the initial value of Toomre's $Q$\@.  We chose to set $Q_{0}$ to a
marginally stable $Q_{0} = 1.1$ over the entire disk. The offset
runs
lasted just under 10 rotation periods, as measured at the
half-mass radius of the
disk. The particles initially form a 
counter-clockwise rotating disk.  

The simulation is done on a Cartesian grid with $256 \times 256$
cells.  During the runs with the intruder present, the test
galaxy and
its halo had a diameter of 128 cells.  To maximize use of
computer
memory and time, the intruder galaxy was a point particle, sent into a
coplanar orbit.  The
perigalactic distance between
the two galaxies was chosen so that they would not overlap even
if they both had realistic radii.  This
precludes any dynamical friction between them.  Their minimum
separation was 42 cell widths and the intruder was on the grid
for
less than one rotation period of the test galaxy. 

During the runs without the intruder, the test galaxy's initial
diameter is enlarged to 192 cells.  This enables us to more
accurately
simulate small-scale features.  For a typical disk galaxy of
radius
10~kpc \citep{BinneyTremaine1987} this translates into a cell
width of
roughly 104~pc.  Each rotation period of the disk corresponds to
roughly 250~Myr.

Runs were made with three different halo mass distributions:

\begin{enumerate}
\item The distribution, as described by
       \citet{SellwoodCarlberg1984} and
       \citet{CarlbergFreedman1985}, that creates
      a rotation curve simulating that of an Sc galaxy.

\item The isothermal distribution described in
       \citet{BinneyTremaine1987}.

\item An isothermal distribution plus the gravitational potential
of a supermassive central black hole.

\end{enumerate}

The rotation curves for these potentials are all presented
together in
Figure~\ref{figure1}. 

Tables \ref{tab1}, \ref{tab2}, and \ref{tab3} list the basic properties and
results for these runs.  Column 1 in each of these tables lists the run numbers
referred to throughout the paper.  All runs used an Sc
distribution for the disk.

In $\S 2$ we present the
results of 5 runs in which an intruder galaxy passes the target
galaxy.  In $\S 3$ we present the results of 12 runs in which
the stars and halo initially have different centers of mass,
along with three reference runs with the same mass distributions
and no
initial offsets.  In $\S 4$ we compare these runs to each other,
to previous simulations, and to observations that they may
explain. In $\S 5$ we present our conclusions.

\section{EFFECTS OF A HIGH SPEED GALACTIC COLLISION}

Disk galaxies interacting with neighboring galaxies have lopsided
structures as seen, for example, in Karachentsev 64 and M101.
Unlike simulations of slow-speed interactions that lead to
mergers \citep{HernquistMihos1995, ZaritskyRix1997}, we send a
simulated companion in a high speed, slightly positive-energy,
coplanar
orbit that takes it past the target galaxy and off the grid.  The
intruder is modeled as a point mass with 20\% the total (halo
plus
disk) mass of the target galaxy.  It is initially sent onto the
grid
passing the target galaxy in the same direction as the target
galaxy's
star particles orbit.  The intruder enters at an angle adjusted
so
that, even if it did have a realistic distribution of stars, it would
not
collide with the target galaxy.  This prevents the merger of the
two
galaxies, which we do not seek to simulate.

The first two intruder runs had the
halo simulated by a second $N$-Body component of 10000 particles
with a different $Q$ than
the star particles.  Therefore, both the star and halo particles
responded to the gravitational
influence of the intruder, as well as to each other.  (Halos that
move we denote as dynamic,
while fixed halos are called static.)  One of these runs
(hereafter Run 1) simulated a dynamic
halo of cold matter, with an initial $Q_{0,halo} = 0.3$.  The
other run
(Run 2) simulated a dynamic halo of hot matter, with $Q_{0,halo}
= 5.0$.

Figure \ref{figure2} shows the displacement of the center
of mass motion
of the stars relative to the center of mass of the halo in Run 1.
The halo's center of mass is
always located at (0.0,0.0) in this, and similar, figures. The
maximum separation between the halo and star centers of mass is
three quarters of a cell width
(78 pc).  Although ``noisy'', Figure \ref{figure2} shows
distinct periodic,
cyclic motion as the two centers of mass waltz around each other. 
The
dynamics of this run are complicated by the fact that the cold
halo
particles develop a bar, which persists throughout the $6\case{1}{4}$
rotation
periods of the run.  
     
Figure \ref{figure3} shows the separation in Run 2
between the stellar and halo centers
of mass as.  This is the hot halo
case.  Due to the
greater kinetic energy in its particles, we expect that the halo
center
of mass will be displaced less than the stars in this run.  This
is
what we observe.  As a result, the maximum separation is nearly
two
grid cells or 208 pc.  No bar develops in this halo.

Next we ran three intruder runs with a hydrodynamic simulation
of the halo.  These were done to avoid the effects of dynamic
friction that were observed to
occur between the star and halo particles (each with initially
 different $Q_{0}$) in Runs 1 and 2. 
We used an Sc mass distribution of compressible, shockable,
gravitating gas.  Instead of assigning an initial $Q$, we give the
gas an
initial uniform energy density.  This is equivalent to giving the
gas a temperature profile. 

Figure \ref{figure4} shows the center of mass separations
between the halo and stars for the case with a low initial energy
density gas
corresponding to an initial gas temperature of $T_{0}=0.5$~K as measured
at the
half mass radius.  During the run, denoted Run 3, the maximum
separation between the halo and star centers of
mass reached nearly 3 cell widths or 312~pc.  The result for the
$T_{0}=5.0$~K
run (Run 4) is similar with a maximum separation of nearly 6 cell
widths.
     
The results of the hottest gas run (Run 5), with $T_{0}=5\times10^{5}$ K,
are
especially intriguing.  Figures \ref{figure5} and \ref{figure6}
show the
gas and star distributions throughout this run.  Note that in
this and all other intruder runs, the intruder travels
counterclockwise,
entering the grid at roughly the 5 o'clock position and leaves at
roughly
the 1 o'clock position at time $t=0.98$ rotation periods.  After
leaving the grid, the intruder's
gravitational potential is ignored.   In response to the
intruder, a
bulge is clearly visible at the three o'clock position in both
the gas
and stars at time $0.861$ rotation periods.  Note the similarity
between timestep 4.898 in Figures~\ref{figure5} and
\ref{figure6} and the optical image of Karachentsev\ 64
(Figure~\ref{figure7}).  Karachentsev\ 64 is a pair of interacting
spiral galaxies.

The centers of mass of both components drift upward throughout
this
run.  This is due to the asymmetry created in the spiral arms by
the
intruder's passage.  The more open arm, on the left, carries
angular
momentum downward and to conserve it, the rest of the system
moves
upward.  The relative displacement between the centers of mass in
this
run is shown in Figure \ref{figure8}.  While the dynamics of
the system
greatly complicate the relative center of mass motion, clearly it
still exists.

Because of the differences in behaviors of particles and gas
acting as the halo, these three gas halo runs cannot be compared
directly with the
previous two $N$-body halo runs.  Nevertheless, the behavior of all
five consistently show
an offset between the two centers of mass.  Furthermore, the
``hotter''
the halo, the greater the offset. 

In all of our runs, the center of mass of the stellar disk moves
relative to the center of mass of its halo as a result of the
gravitational interaction with an intruder.  This provides the
motivation to explore the dynamics of visible matter sloshing
around in
a deep gravitational halo.  The difficulties with doing this
following
the perturbation of an intruder, as in Runs 1 through 5, are
two-fold.  First, as we saw in Run
5, the transients created by the intruder's extremely powerful
and
asymmetric passage make it difficult to see how the star system
is
responding just to its halo's potential.  Second, there may be
other
mechanisms that cause a separation of centers of mass that are
not so
disturbing to the system.  Therefore, we now look at systems with
initial offsets between the star and halo centers of mass, but
without
any intruder to mix up the system.

\section{OFFSET RUNS}

\subsection{Dynamic Halo Runs}

We proceed now by displacing a variety of stellar disks from
the centers of mass of their halos. Our first four offset runs, labeled
Runs 6 -- 9, have
both stars and halo free to move.  Predictably, the two runs with
dynamic $N$-body halos show
significant dynamical friction between the stars and halo
particles.  These lead to rapid damping
of the center of mass motion.  In the low angular momentum Run 6,
the stars and halo have
essentially the same centers of mass and are in equilibrium
configurations within less than four rotation
periods.  The relative positions of the centers of mass show
little, if any, periodic motion during the run (Figure \ref{figure9}).

Transient structure occurs in the stars in Run 6.  They evolve
a multiarmed spiral psttern which changes to a
two-armed spiral and then to a symmetric disk by the beginning
of the fourth rotation period. 
This is also a typical evolution of a non-offset $N$-body system moving in the
potential of a fixed halo.  In
comparison, the transient structures seen in the high angular
momentum version, Run 7, shows
a
significant feathery one-armed spiral structure that evolves to a
small bar with a ring surrounding
it and then to a symmetric system, after less than
four rotation periods (Figure~\ref{figure10}).  Similar one-arm features
have been reported by \citet{LotanLuban1990}, as mentioned in
\citet{Struck1999}, in head-on, low impact parameter
collisions.  The center of mass of the
stars in our Run 7 follow an arc
inward, dampened after 1.8
rotation periods by the dynamical friction, again with little
indication of periodic motion relative
to the halo's center of mass.

If, as seems likely, most of the true halo mass is a smoother
distribution of matter, less prone to the
effects of dynamical friction, it would be more appropriate to
simulate it's dynamic behavior by that of a gravitating fluid.  Therefore,
we made two runs with our gravitating gas component serving as
the halo, as in Runs 3 -- 5,
above.  Run 8 has the same initial conditions as Run 6: hot halo,
large initial center of mass
offset, and small initial angular momentum for the stars. 

Both the halo gas and the stars in Run 8 go
through non-axisymmetric transients.  After about $1\case{3}{4}$
rotation periods, the gas settles into a
two-armed spiral structure.  The stars develop a distinct bar
with spiral arms winding out from its
ends.  The spiral structure is much more distinct in the gas than
in the star particle component
(Figures \ref{figure11a} and \ref{figure11b}).  

Unlike the runs with dynamical friction, the center of mass of
the stars for the present run show
distinct orbital motion around the center of halo mass (Figure
\ref{figure12a}).  The halo's center of mass
moved less than a grid cell width during any fifty timesteps,
during which the star's center of
mass moved up to three cell widths (Figure \ref{figure12b}).
This is important in our justifying
shortly the value of examining runs
with fixed halos.

Run 9 is the high angular momentum version of Run 8.  In all
other respects the initial conditions
of the two runs are the same.  The qualitative evolution of the
gas is similar in both, although the
concentration of gas in the center of the system is lower in the
higher angular momentum case.  
As a result, the halo in Run 9 remains axisymmetric over a larger
radius.  The initial perturbation on
the stars due to the halo is lower in this run than in Run 8, because the stars
are not plunging toward the center
of the halo's mass distribution.  The stars develop transient
arms, but the bar is much less
concentrated than in Run 8.  Indeed, the central region of the
stars remains axisymmetric (Figure~\ref{figure13b}).
The higher angular momentum of the stars gives the halo
more time to move during each
timestep. The
stellar center of mass shows a more
circular orbit around the halo center of mass than in Run 8
(Figure \ref{figure14}).  This is consistent with
the decreased perturbation caused by the stars on the halo.

The orbital motion of the stellar component's center of mass
around the halo seen in the last two
runs and the associated structure in the star component suggest
that such sloshing will have
observable effects on galaxies.  This belief is strengthened when
we compare these results to a
``traditional" run, in which the centers of mass of the disk and
halo are initially the
same and remain the same throughout the run.  Labeled Run 10 in
Table \ref{tab3}, the non-offset Sc disk shows the typical evolution seen
in $N$-body simulations of the disk.  It goes through an initial
high-m
(multiarmed) spiral phase, and then into a distribution with a
persistent bar, without spiral arms (Figure \ref{figure15}).  This bar's long
axis extends one quarter of the
diameter of the disk.  
     
Run 10 shows profound differences in mass distribution compared
to Run 8, which also develops
a bar.  While the non-offset star distribution maintains an
overall disk-shaped structure,
virtually all the star particles in Run 8 are either in the bar
or in the two weak spiral arms, while there are many stars in an axisymmetric distribution
surround the bar throughout Run 10. 
Furthermore, the bar in Run 8 is much thicker than the bar in Run
10.  Although both runs have
bars, clearly, the stars in the run with the dynamic halo (Run 8)
have followed a different
dynamical evolution than the run with the static halo.

The underlying spiral structure created in the halo of these runs
is of concern, as there is no
observational
evidence yet for such structure.  To eliminate any
nonaxisymmetric effects from a halo, we have
made one final set of runs in which the halo remains fixed and
axisymmetric.  This construct will
affect the timescales for the motion of the stellar center of
mass around the halo center of mass. 
However, as noted above, the relatively small motions of the halo
in the low angular momentum
situation suggest that this effect will be relatively small for
these runs.   Indeed, since the halos
are expected to be even higher than the 75\% of the total
galactic mass we have used here, the
effect of halo motion will be even less than Runs 8 and 9
indicate.  Furthermore, the savings in
computer time enabled us to make many more runs with fixed halos
than we could have
otherwise accomplished.

Runs 11 through 16, 18, and 20 (see Table \ref{tab3}) have displaced stellar centers of mass in fixed halo
mass
distributions. (Runs 10, 17, and 19 are undisplaced reference runs.)  As seen in
Table \ref{tab3}, the four parameters we studied can be summarized as
follows:

\begin{itemize}
\item Initial Offset (Column 2):  The initial offset of the
disk's
center of mass is either $2 \sqrt(2)$ cells (hereafter, ``small
offset") or $10 \sqrt(2)$ cells (``large offset") from the halo's
(fixed) center of mass.  The small offset runs correspond to
3\% of
the disk's radius.  A grid cell corresponds to roughly 104~pc.
\item Angular Momentum (Column 3):  The disk's center of mass
initially
has either a low angular momentum or a high angular momentum
around the
halo center of mass.  The high angular momentum runs' angular
momenta
are 6\% higher than the low angular momentum runs.
\item Mass Distribution (Column 4):  The halo has a variety of
radial mass distributions, as described above.  The disk is
always an
Sc distribution.
\item Counter-rotating Component (Column 5): In some of the runs
we
have set 50\% of the star particles in counter-rotating motion.
\end{itemize}

\subsection{Sc Halo Mass Distribution}

Run 11 is a small offset, low center of mass angular momentum run.  This angular
momentum is in the same direction as the stellar angular
momentum.  As with all the off-center runs reported in this
paper, this run shows an initial transient containing more spiral
structure than the traditional run (Run 10).  In this offset run,
the
transient is followed by a period of spiral structure with eight
or
nine arms that combine and decrease in number.  This is
followed by the establishment of a bar.  The bar becomes robust
and permanent after 4 rotation periods. The amounts of $m=1$
(one-armed spiral), and $m=2$ (bar) motion are comparable to the
traditional run.  This evolution is summarized in
Figures~\ref{figure16}
and \ref{figure17}.  The center of mass motion of this system
spirals
around the halo center of mass (Figure~\ref{figure18}), as also seen
in
\citet{LevineSparke1998}.

Run 12 has an initial offset 5 times larger than Run 11, but is
otherwise the same.  There is a marked transient one-armed
feature that dominates the structure for about $\case{3}{4}$ of a rotation
period.  This is followed by several rotation periods in which
multiple, flocculent arms spiral out from the m=1 arm.  The m=1
feature persists for the entire run, long after the higher-m
structure has dissipated.  By the end of 10 rotation periods,
there is still no sign of a bar.  The center of mass of the disk
in this run spirals inward less than is seen in Run 11
(Figure~\ref{figure19}).

Run 13 is a high angular momentum version of Run 12.  The initial
one-armed spiral is 50\% stronger than in Run 12. 
Particularly
notable about this run is the formation of several very high
density concentrations of $N-$body particles (Figure~\ref{figure20}). 
These regions persist for about $1\case{1}{2}$ rotation periods.  The end
state
of the run after 10 rotation periods is qualitatively the same as
in
Run 12. No bar develops during this run.

The center of mass of Run 13 again spirals in toward the center
of
the halo mass, but the higher angular momentum of this system
results in a longer time during which the two centers of mass
are different. The results of this run are consistent with a
similar run presented by \citet{LevineSparke1998}.

Run 14 is the same as Run 13 except that the initial angular
momentum of the disk's center of mass is in the opposite sense to
the particles' orbital motion. The dynamics of the resulting
one-armed spiral are significantly different than in Run 13.  In
the
present run, the spiral is stationary in angular location for
nearly 1.5 rotation periods, while continually expanding radially
(Figures~\ref{figure21a}, \ref{figure21b}, and \ref{figure21c}). The arm then
transforms into a wide, diffuse, leading arm spiral that persists
for
the remainder of the run.  No bar develops during this run.

Recalling the leading arm spirals we saw develop in runs with
counter-rotating disks \citep{Cominsetal1997}, we made two Sc
runs with
counter-rotating disk particles. Both of these runs had low
angular
momentum orbits for the disk's center of mass.

Run 15 has equal numbers of particles rotating in opposite
directions, but is otherwise the same as Run 11.  Each of the two
counter-rotating sets of particles were separately given the same
low angular momentum, in opposite directions.  As a result, of
course, the total angular momentum of the system was zero.  The
initial offset served as a perturbation that created tightly
wound spirals in the inner region of the disk
(Figure~\ref{figure22a}).
These evolved after $1\case{1}{2}$ rotation periods into the same
single-armed
spiral that we saw in \citet{Cominsetal1997}. This spiral
expanded
radially (Figure~\ref{figure22b}) as far out as the inner Lindblad
resonance where its outer edge bifurcated into spirals like the
tongue
of a snake (Figure~\ref{figure22c}). At this point the spiral arm is
about
to change directions. While damping thereafter, the spiral
structure
persisted for the remainder of the run.  The center of mass of
the
entire disk followed a damped oscillation across the halo center
of
mass throughout the run.

Run 16 also had half of its particles counter-rotating, but
otherwise is the same as Run 12.  The center of mass of its disk
is displaced 5 times farther from the halo center of mass than in
Run 15.  Again, the center of mass of the entire disk followed a
damped oscillation across the halo center of mass, while the
individual counter-rotating halves of the disk separately
spiraled
inward.

The combined motion created by the initial center of mass offset
and the counter-rotating disks led to significantly different
behavior than any other run.  Within one quarter of a rotation
period, the system had developed a reflection-symmetric pattern
(Figure~\ref{figure23}).  After three more rotation periods, a
small
central bar formed, surrounded by a ring, from which a single,
tightly-wound arm spiraled.  Thereafter, the bar oscillated in
and out
of existence, due to energy transfer from the bar to the spiral or
ring structure.  The spiral arm decayed over 6 rotation periods,
reversing its direction several times as it
decayed.

\subsection{Isothermal Halo Mass Distribution}

In order to learn how robust these results are, we then made a
series of runs using different halo mass distributions, while
keeping the Sc particle disk. In this section we present the
runs using an isothermal halo. 

Run 17 is a reference case without initial offset. Comparing the
results
for this case   to the comparable Sc case (Run 10), we note that
Run 17
develops a bar more quickly, but that this bar is less elliptical
than
the one in Run 10, extending only over 15\% of the disk's
diameter.
This bar persists for the entire run.
The spiral structure in the two runs (Run 10 and Run 17) are
similar in
numbers of arms and evolution. While specific differences between
the
particle dynamics are perceptible between the two runs, they are
subjectively very similar throughout.

Run 18 has a high offset disk and low angular momentum. This
isothermal run is analogous to Run 12, above. The global spiral
features of these two runs are essentially the same. After the
transient spiral stage, however, Run 18 develops a short bar,
unlike Run 12. The bar extends over 15\% of the disk's diameter
by the fifth rotation period, despite the fact that its $Q$ is by
then over 4.0, and climbing. The bar's m=2 Fourier amplitude is
20\% less than that of the bar in Run 17.

\subsection{Isothermal Halo Mass Distribution With Galactic Black Hole}

Run 19 adds the gravitational influence of a supermassive black
hole at the halo's center of mass to Run 17.  The black hole's
mass is
about $4.2\times 10^{8}M_{\odot}$, which corresponds to 0.5\% of
the
mass of our model galaxy. The disk in this run
has no initial offset. The black hole has the effect of
suppressing the bar, while allowing virtually the same transient
spiral structure to unfold. The run ends with a high particle
concentration near the center of the disk, consistent with the
added gravitational potential of the black hole.

Run 20 adds a large offset and high angular momentum to Run 19.
Run 20 is therefore analogous to Run 13 (Sc disk and halo). The
addition
of the black hole mass is significant, especially during the
first
three rotation periods. Specifically, we observe richer
concentrations
of particles in Run 20 than in any other run
(Figure~\ref{figure24}). A
bar forms in this run. Its end state is very similar to that of
Run 18,
differing only in the greater central concentration of particles
in the
run with the black hole.

\section{ANALYSIS AND DISCUSSION}

\subsection{Theory}

\citet{Lovelaceetal1999} studied the  eccentric dynamics of a
disk with
an exponential surface density distribution [$\Sigma_d
\propto \exp(-r/r_d)$] represented by a large number of rings and
a
central mass $M_0 \sim 10^9M_\odot$ embedded in a passive dark
matter
halo. 
   The inner part of the disk $r\lesssim 2.5$ kpc was found to be
strongly unstable with $e-$folding time $\sim 30$ Myr for the
conditions considered.
   Angular momentum of the rings is transferred outward, and to
the central mass if it is  present.
   A trailing one-armed spiral wave is formed in the disk.
   This differs from the prediction of \citet{Baldwinetal1980} of
a
leading one-armed spiral.
    The outer part of the disk $r \gtrsim r_d$ is stable and in
this
region the angular momentum is transported by the wave.
    This instability appears qualitatively similar to that found
by \citet{TagaIye1998b} for a fluid Kuzmin disk with surface
density
$\Sigma_d \propto 1/(1+r^2)^{3/2}$ with a point mass at the
center
where unstable trailing one-armed spiral waves are found.
    The present findings do not give evidence of instability of
the
inner part of the disk, but they do indicate that a long-lasting
trailing spiral wave is generated in the disk as found by
\citet{Lovelaceetal1999} and \citet{TagaIye1998b}.

\subsection{Effects of Dynamic Halos}

Particle and hydrodynamic simulations of a galaxy's halo lead to significantly different histories
for the star particles.  This is due in large measure to the dynamical friction that occurs between
halo and star particles.  Any displacement between the
particle halo and stellar centers of mass is quickly
damped.  The hydrodynamic halo does not damp the stellar motion nearly as much, allowing for
prolonged sloshing of the stars and concomitant nonaxisymmetric evolution.   

All other things being equal, lower temperature (or lower energy) halo mass leads to more halo
displacement as the result of perturbation by a passing galaxy or sloshing stars. These
simulations do raise the question as to whether nonaxisymmetric perturbations of the visible
matter in galaxies is accompanied by similar changes in the distribution of the underlying halo.


\subsection{Results of Perturbed Static Halo Simulations}

\subsubsection{Effects of Stellar Disk Displacement from the Halo Center of
Mass and of Disk Angular Momentum on Bar Formation}

The runs presented
here indicate that bar suppression can be caused by energy
redistribution due to the transient motion of the disk center of mass moving
toward the halo center of mass. In general, a larger offset of the galactic disk from the halo
center
is more effective in suppressing the formation of a central bar.
The exceptions to this rule are the runs including counter-rotating
components (Run 15 and Run 16), where in fact the opposite is
observed: 
the smaller offset produces no bar, but the larger offset does.
\citet{Cominsetal1997} showed that the presence of undisplaced 50\%
counter-rotating components is already an effective suppressor of
the
formation of a central bar.

The initial offset of the stellar disk's center of mass 
was modelled as having occurred in
one of two ways: for some runs the stellar disk's center of mass was given
nearly zero
initial
velocity, causing the center of mass to have a very small angular
momentum with respect to the halo's center.
Therefore the stellar disk
followed a
trajectory that leads nearly through the halo center.  In other runs
the displaced stellar disk's center of
mass was given a small ``push" perpendicular to its offset.  This caused
the
 disk to follow a spiral orbit around the halo
center. Our results indicate that the amount of the initial
displacement of the galactic disk has a much stronger influence
on the
evolution of the model galaxy than the amount of its angular
momentum. 
In fact, Runs 12 and 13 have identical initial parameters,
differing only
by their initial angular momentum, and they show no qualitative
difference
in star distribution
by the end of the runs after 10 rotation periods.

\subsubsection{Effects of Different Halo Potentials}

We found little difference in the effect on bar formation from using
different halo mass distributions except for 
the isothermal potential without a central black hole.  Here a
bar
forms for both a small and a large initial offset. This occurred because
the potential of the isothermal mass distribution is about 25\%
lower in
the center of the simulation galaxy, compared to the potential of
the
other mass distributions at the center of the galaxy.  This
reduced
potential at the galaxy's center is the result of our effort to
fit the
isothermal potential closely to the Sc potential over the widest
possible range outside the center.

In the case of the isothermal potential with a central black
hole, the
mass of the black hole was chosen such that its gravitational
potential
combined with the isothermal potential of the halo produces a
potential
that is fitted closely to the Sc potential over the entire range
of the
simulation galaxy, including its center.  A mass of $4.2\times
10^{8}M_{\odot}$ for
the black hole achieved this.

\subsubsection{Effects of Counter-rotating Stars}

\citet{Cominsetal1997} showed that the presence of
counter-rotating
components in the galactic disk can
suppress
the formation of a central bar. Those runs were performed without
any
offset of the stellar components, and their results were similar
to our
Run 15, which had a small offset and exhibited no sign of a bar
at any
stage of the run.  In contrast to this observation, a large
offset as
in Run 16 produces an alternating sequence of a central bar and
an
axisymmetric central feature. This is due to the superposition of
two
motions:  the rotation of the bar, and the back and forth motion
of the
disk's center of mass as described above.  At its greatest
length, the
bar extends close to the corotation resonance.

\acknowledgements{T. Z. and N. F. C. wish to thank
Sun Microsystems, Inc., and EDS Data Systems for their generous
donations of computers on which
most of this work was done.  The work
of R.~V.~E.~L. was supported
in part by NSF grant AST~93-20068.}

\figcaption[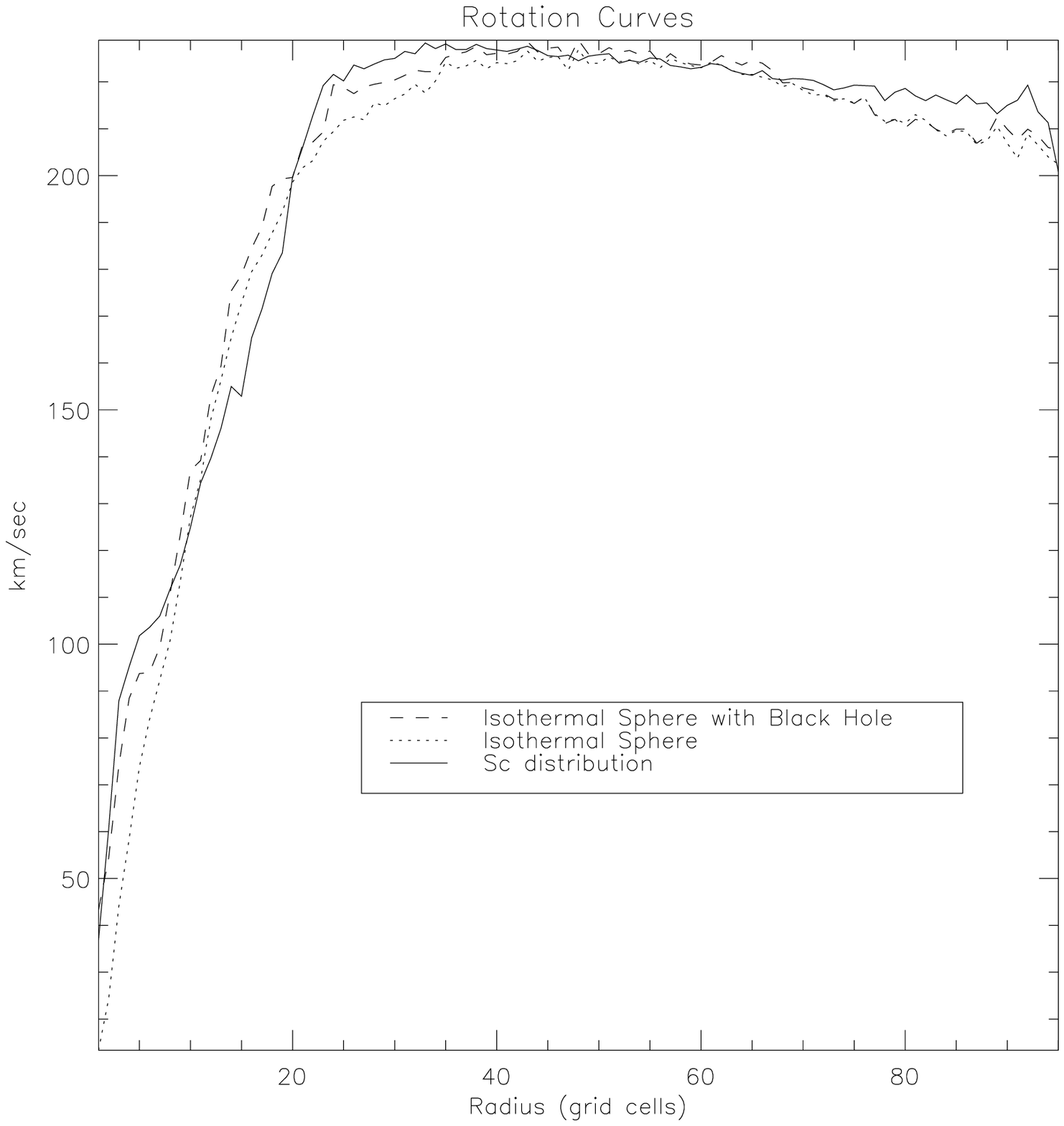]{Rotation curves for the Sc, isothermal,
and
isothermal plus black hole potentials.
The fraction of mass in the
halo was chosen so that the
isothermal potential stabilizes
against the formation of
a bar. An Sc potential
with the same total mass
fraction and with no initial
disk offset does allow a bar to
develop. The radius is given in grid cells,
where one grid cell corresponds to about 104~pc.\label{figure1}}

\figcaption[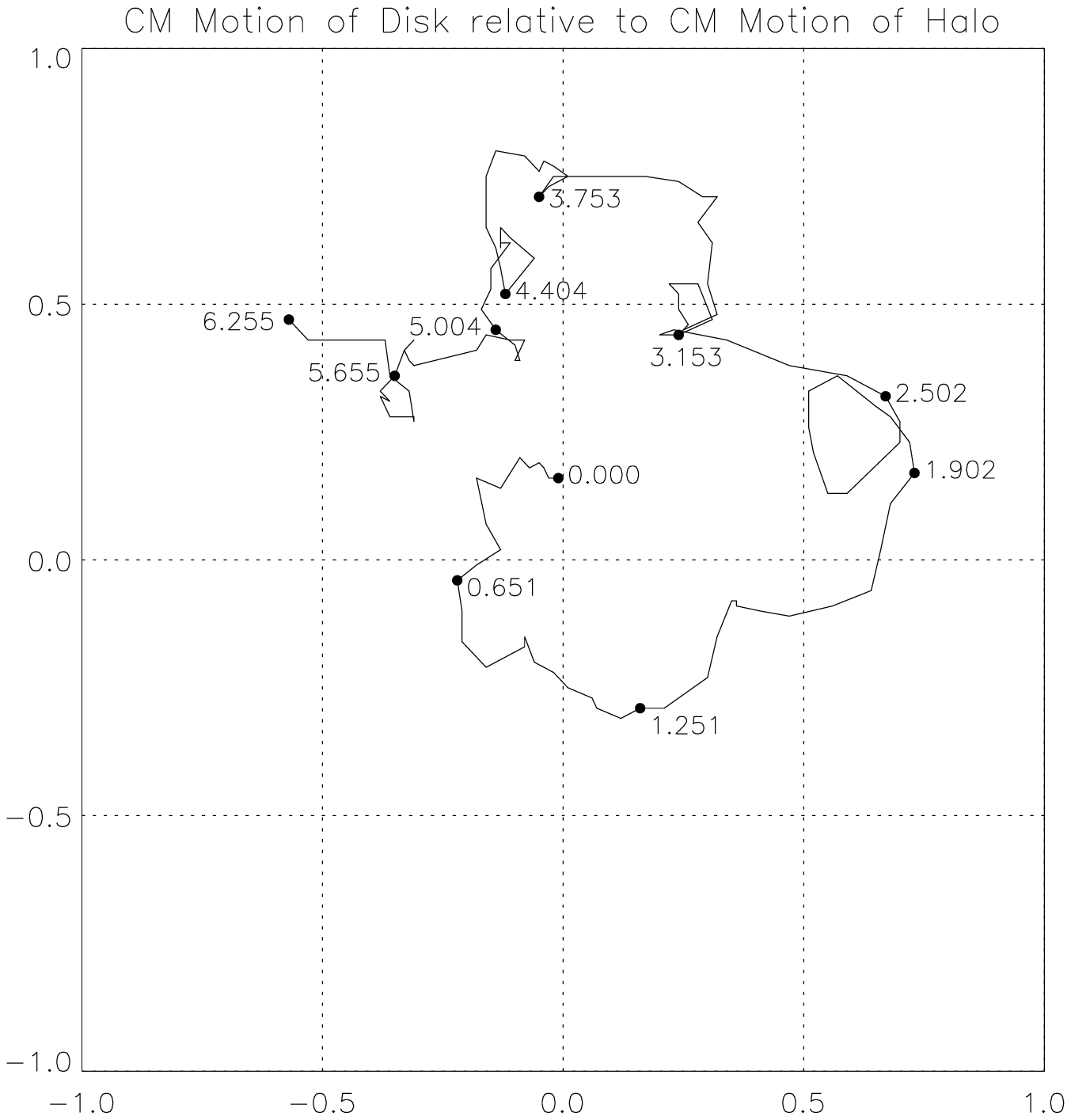]{The motion of the disk's
center of mass, relative to the motion of the halo's center of
mass,
for the cold halo case comprised of $N$-body components.  The
motion starts
at (0,0.2) when the two components almost coincide at the
beginning
of the simulation.  The disk's center of mass then  ``dances''
counterclockwise around the halo's center of
mass.  The numbers along the path are time units in revolutions.\label{figure2}}

\figcaption[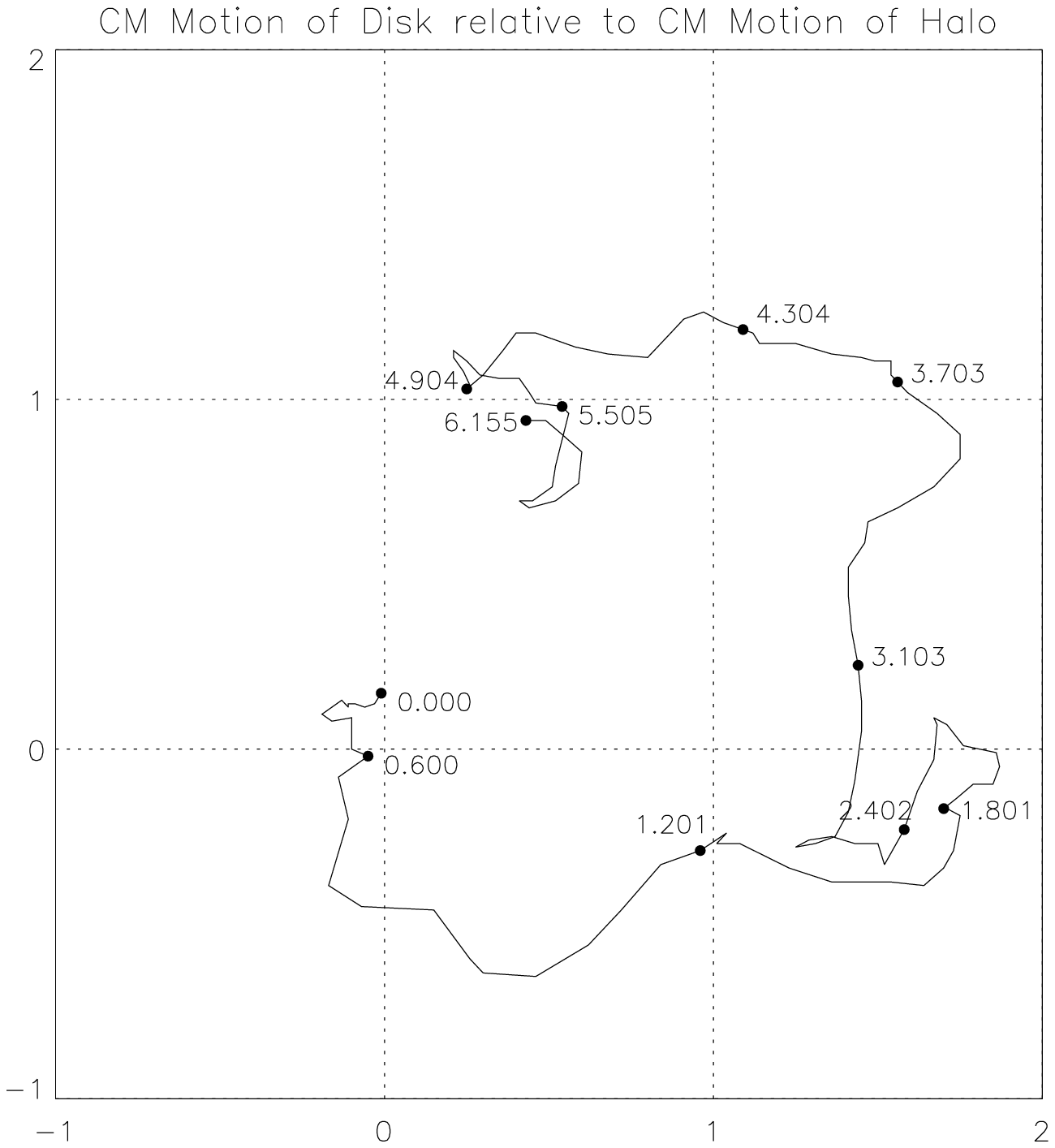]{Same as figure \ref{figure2},
but
for the hot halo $N$-body case.  The beginning of the motion is
again at point (0,0.2).
The numbers along the path are time units in revolutions.\label{figure3}}

\figcaption[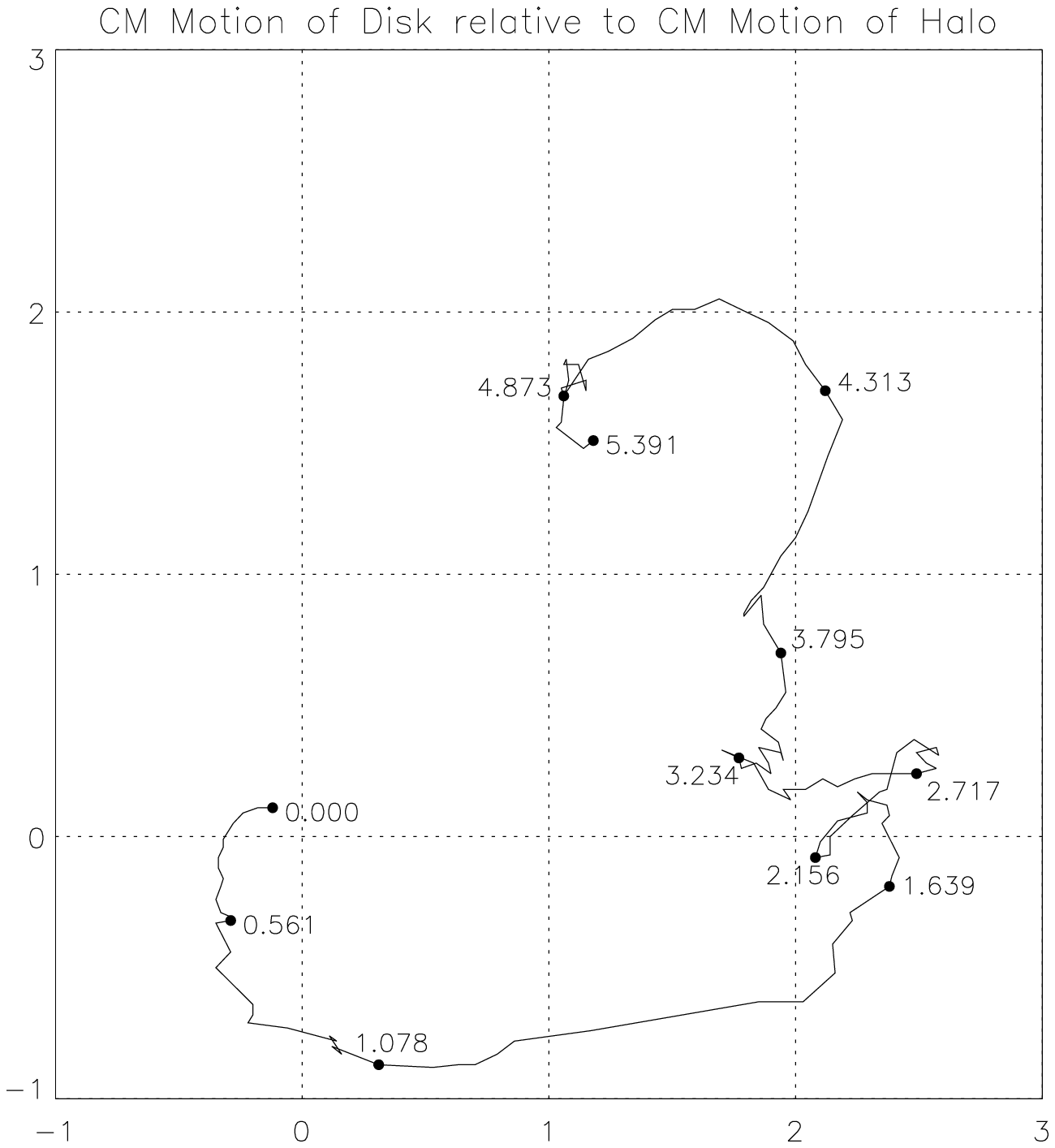]{The center of mass separation between
the halo
and the stars for the cold gas case.  The motion starts at
(-0.1,0.1).
The numbers along the path are time units in revolutions.
\label{figure4}}

\figcaption[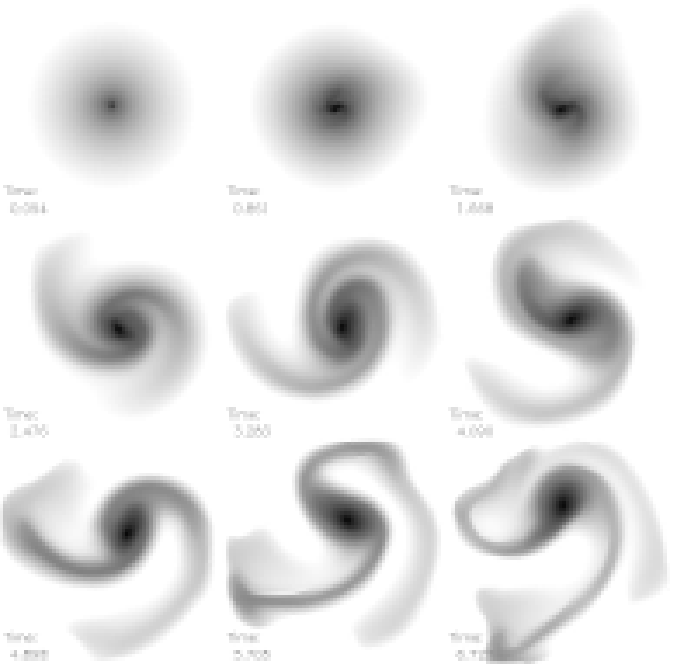]{Gas distribution for Run
5.\label{figure5}}

\figcaption[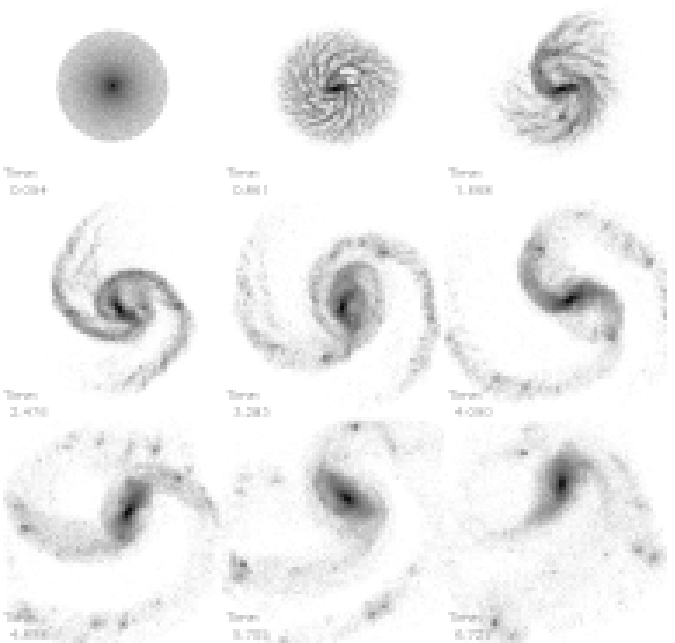]{Star distribution for Run
5.\label{figure6}}

\figcaption[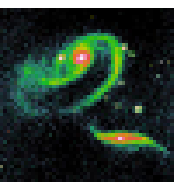]{Karachentsev 64, a pair of interacting
galaxies in Andromeda.  Photo Credit: NOAO/NSF\label{figure7}}

\figcaption[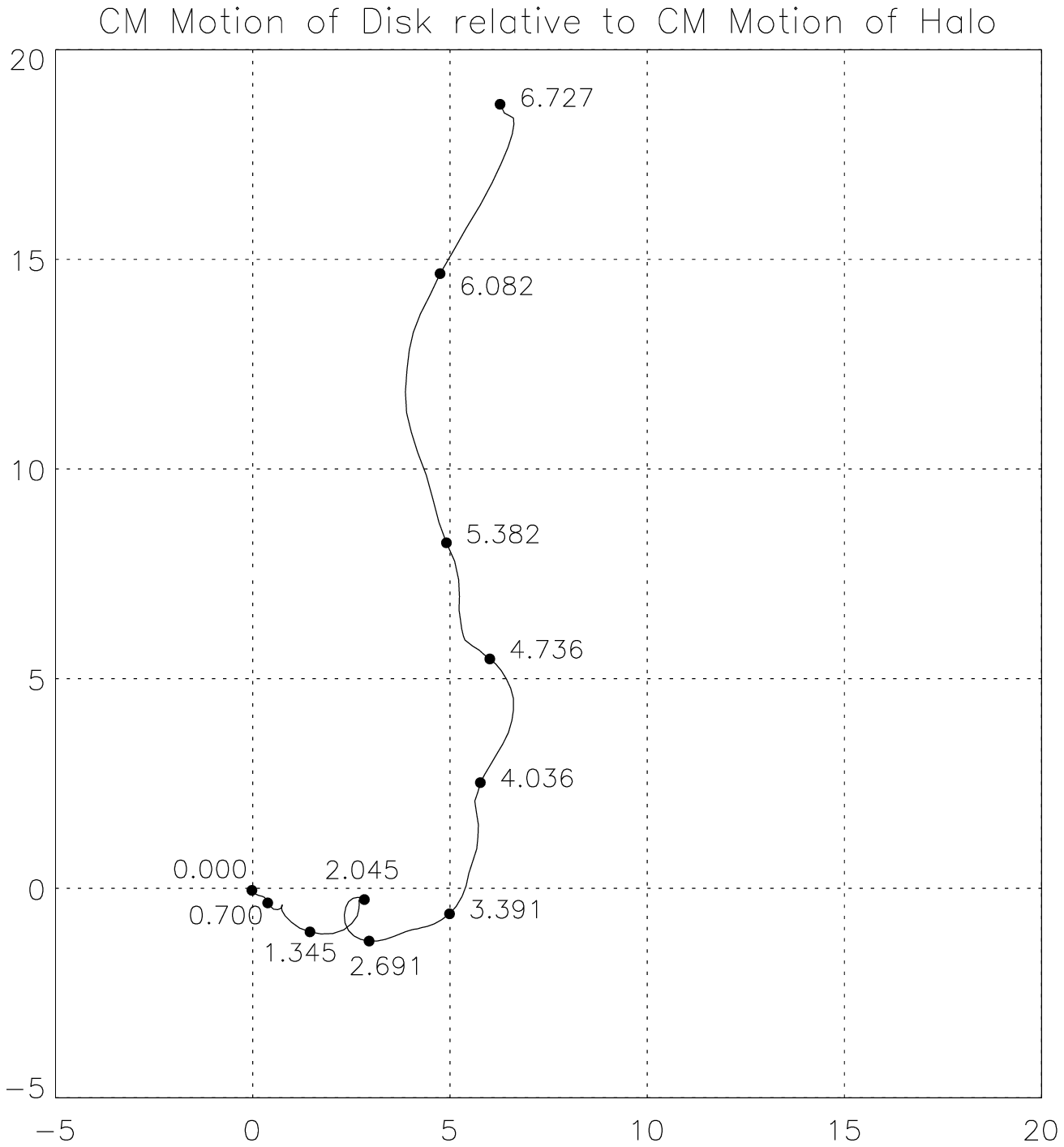]{The
center of mass separation between
the halo
and the stars for the hot
gas case with intruder (Run 5).\label{figure8}}

\figcaption[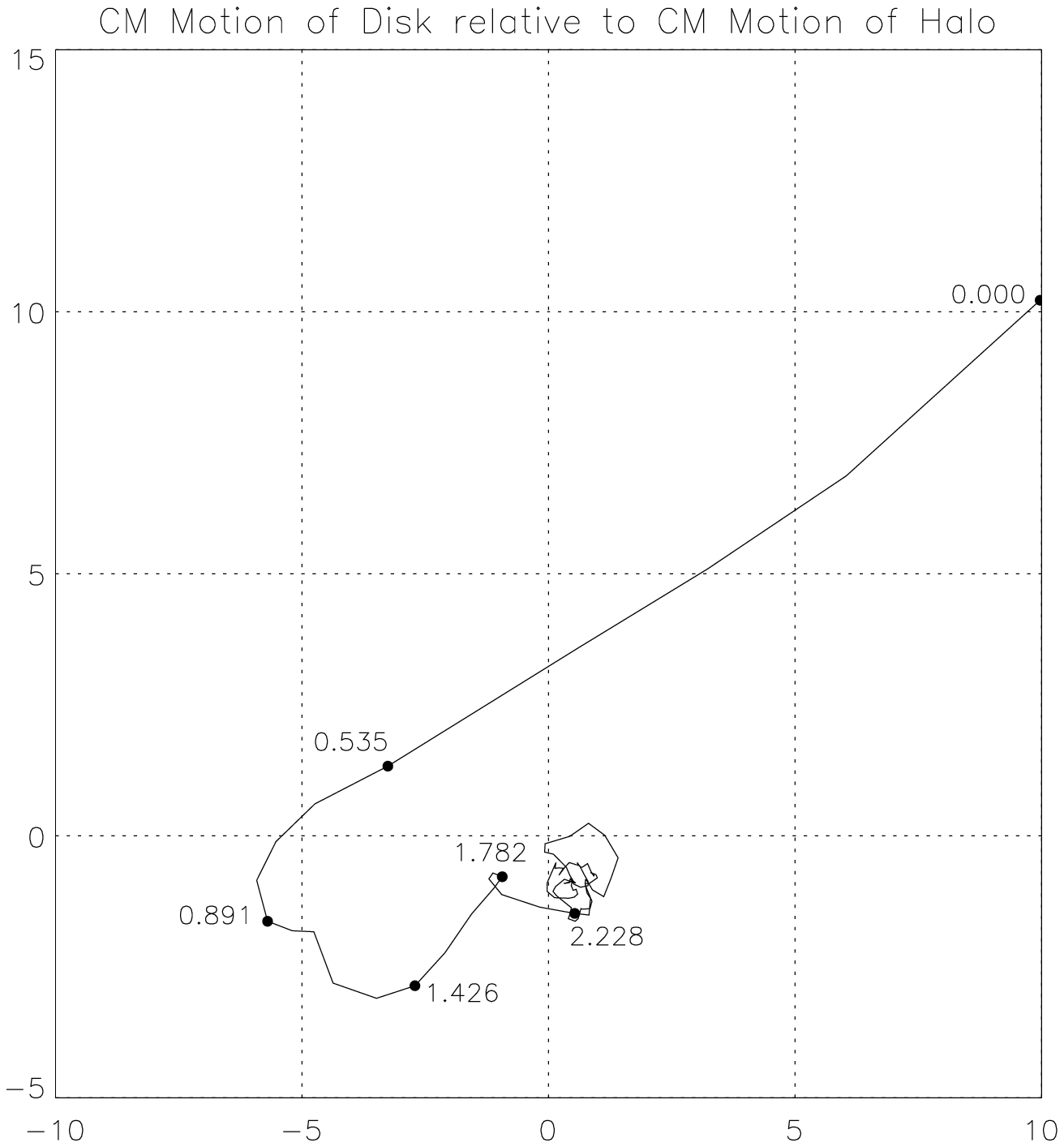]{The center of mass separation between
the halo and the stars for Run 6.  After about $2.5$ rotation periods
the two centers of mass stay very close together for the remainder of
the run, and therefore no further time indicators are included
in this plot.\label{figure9}}

\figcaption[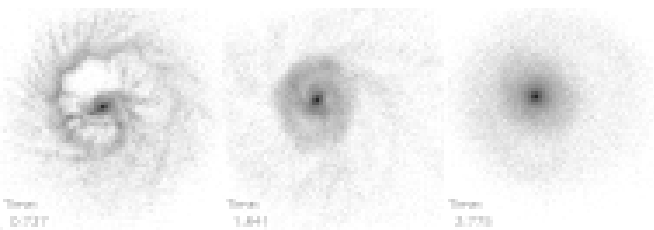]{The stellar density distribution of Run 7 at three
different time steps.\label{figure10}}

\figcaption[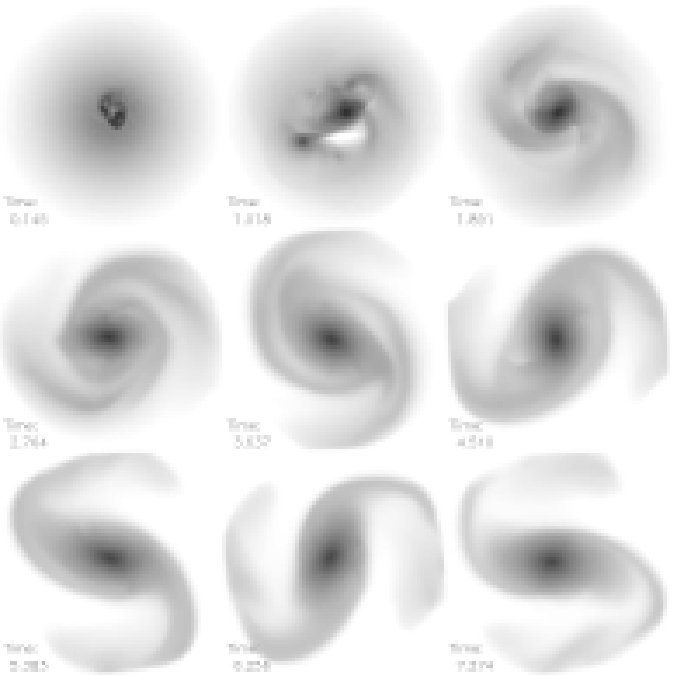,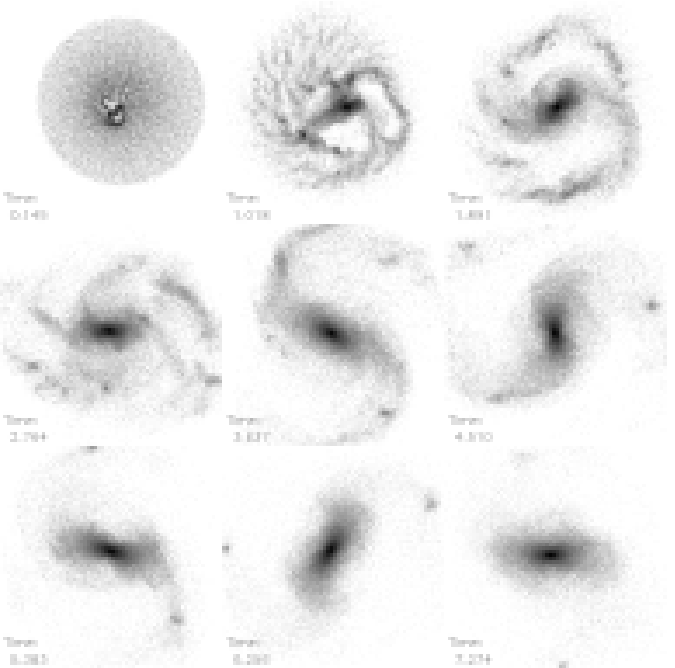]{Run 8: (a) Hot gas distribution, (b) 
star distribution.\label{figure11}}

\figcaption[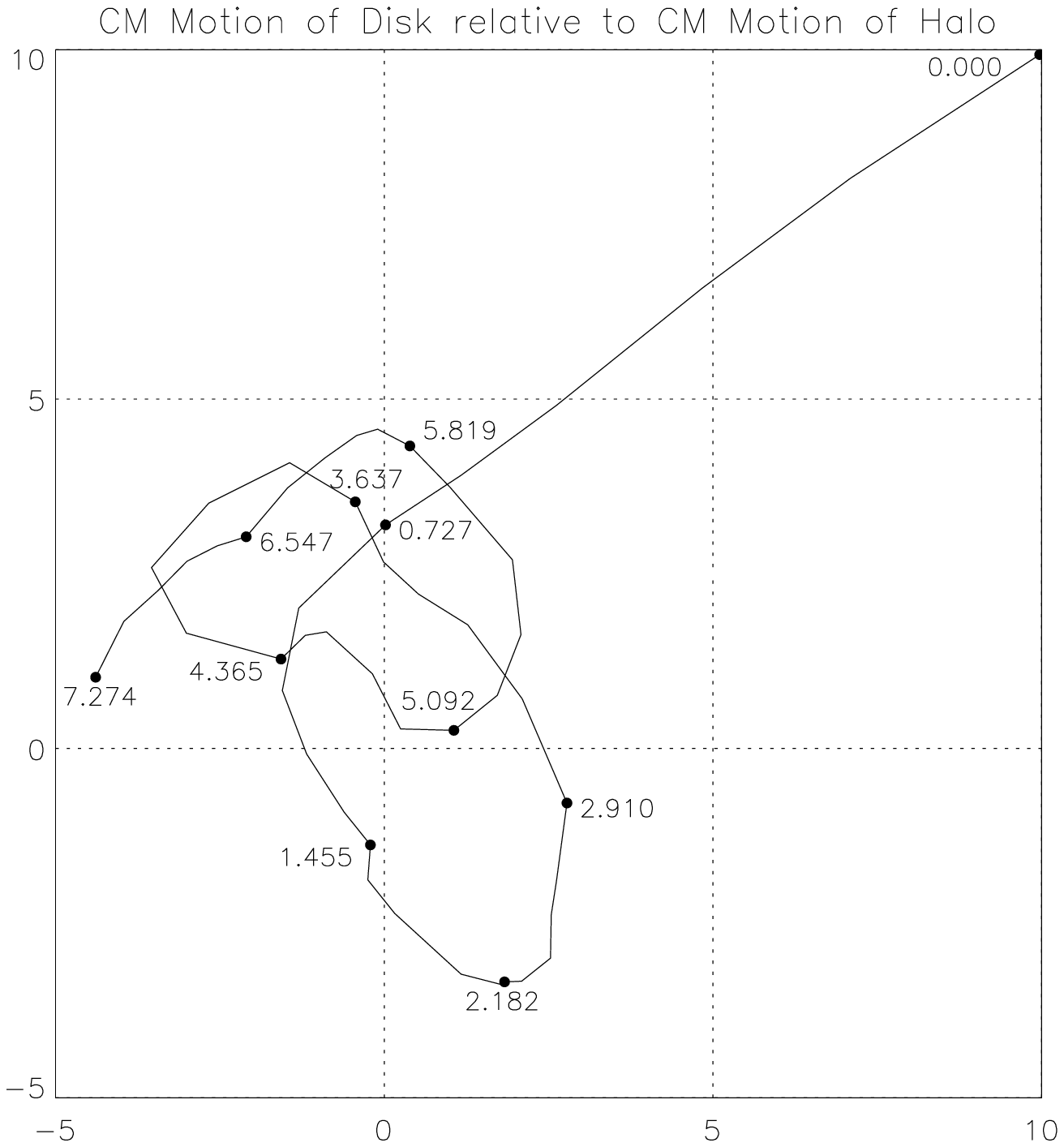,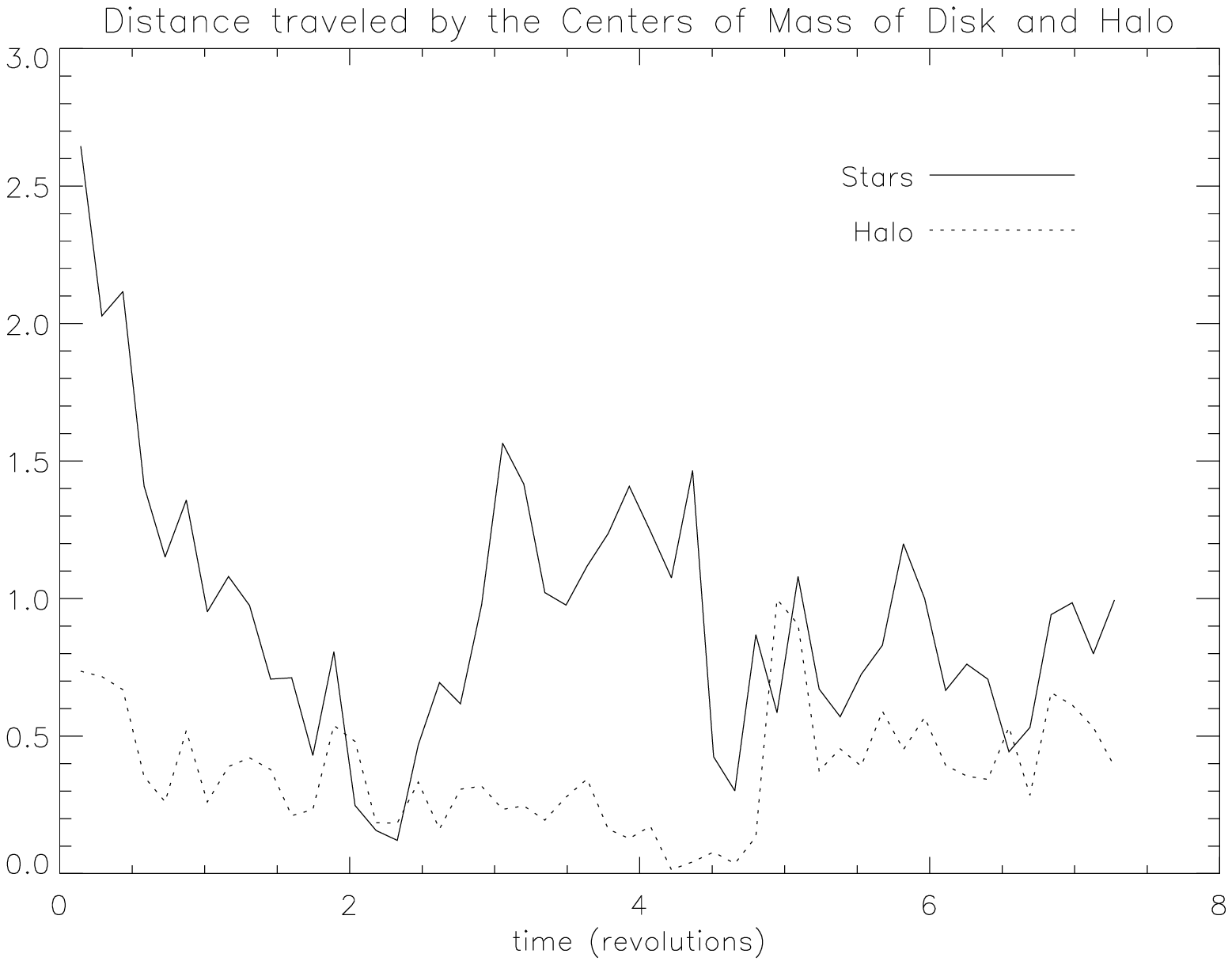]{Run 8: (a) The center of mass separation
between the halo and the stars.
(b) The distances traveled by the centers of mass of the stars
and the halo, respectively.\label{figure12}}

\figcaption[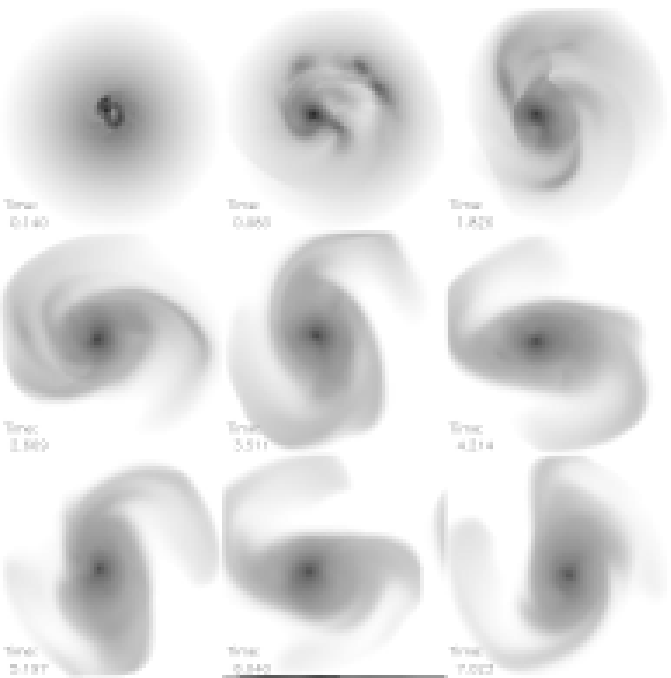,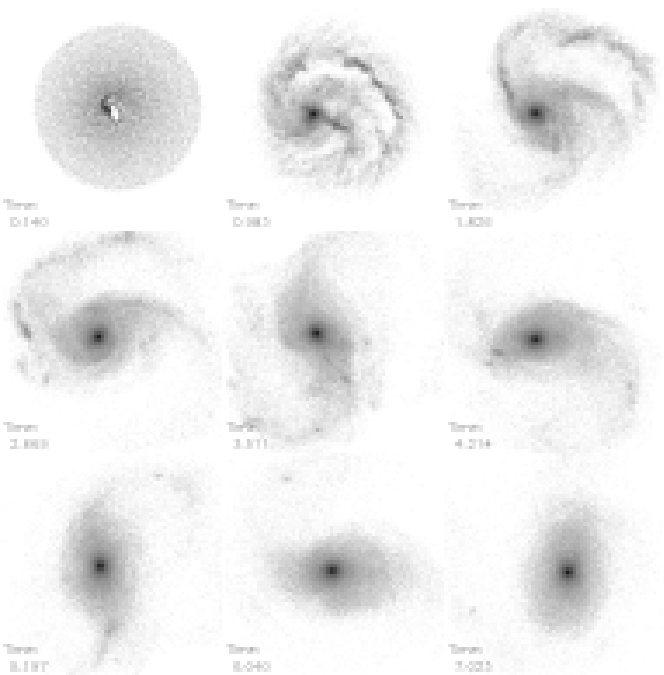]{Run 9: (a) Hot gas distribution (the image
at $t=6.040$ revolutions is corrupted), (b) 
star distribution.\label{figure13}}

\figcaption[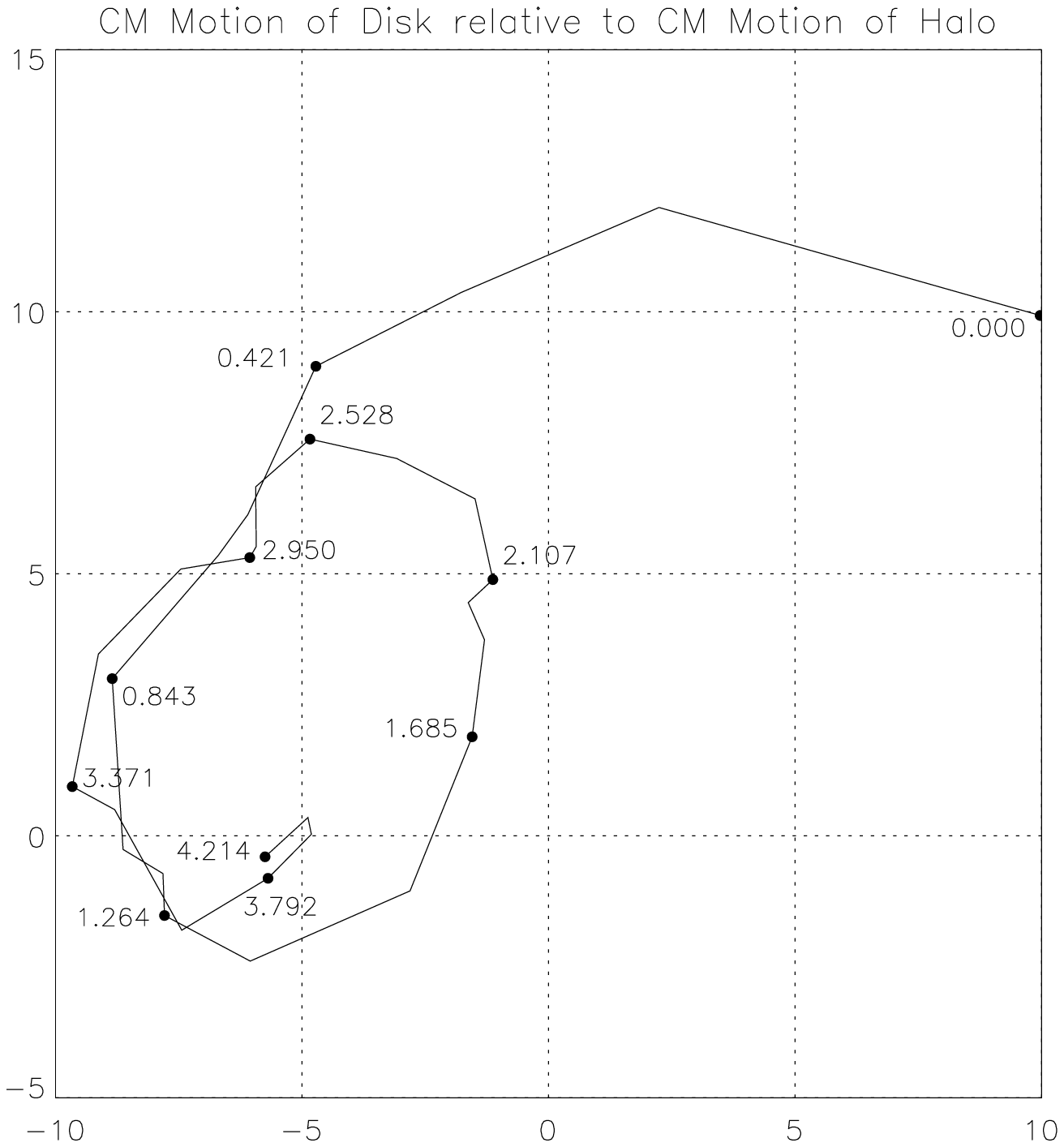]{Run 9: The center of mass separation between the halo and
the stars.\label{figure14}}

\figcaption[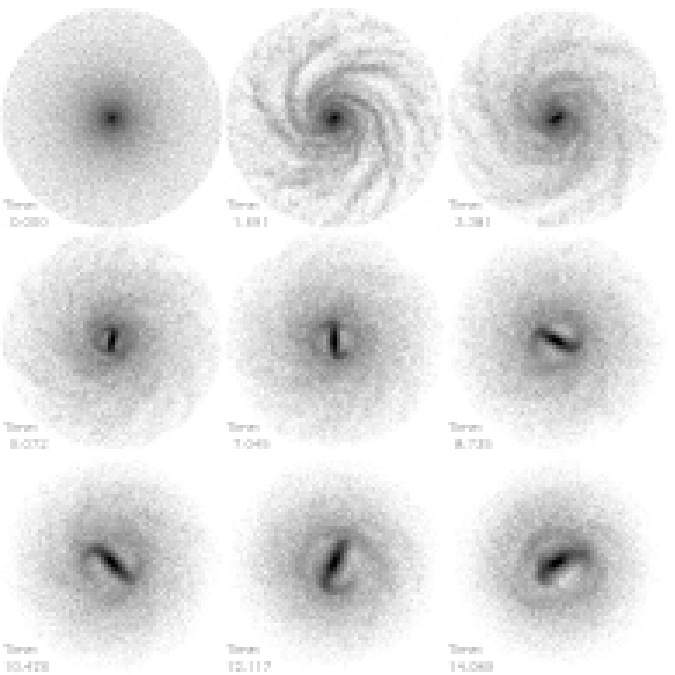]{Run 10: Star distribution.\label{figure15}}

\figcaption[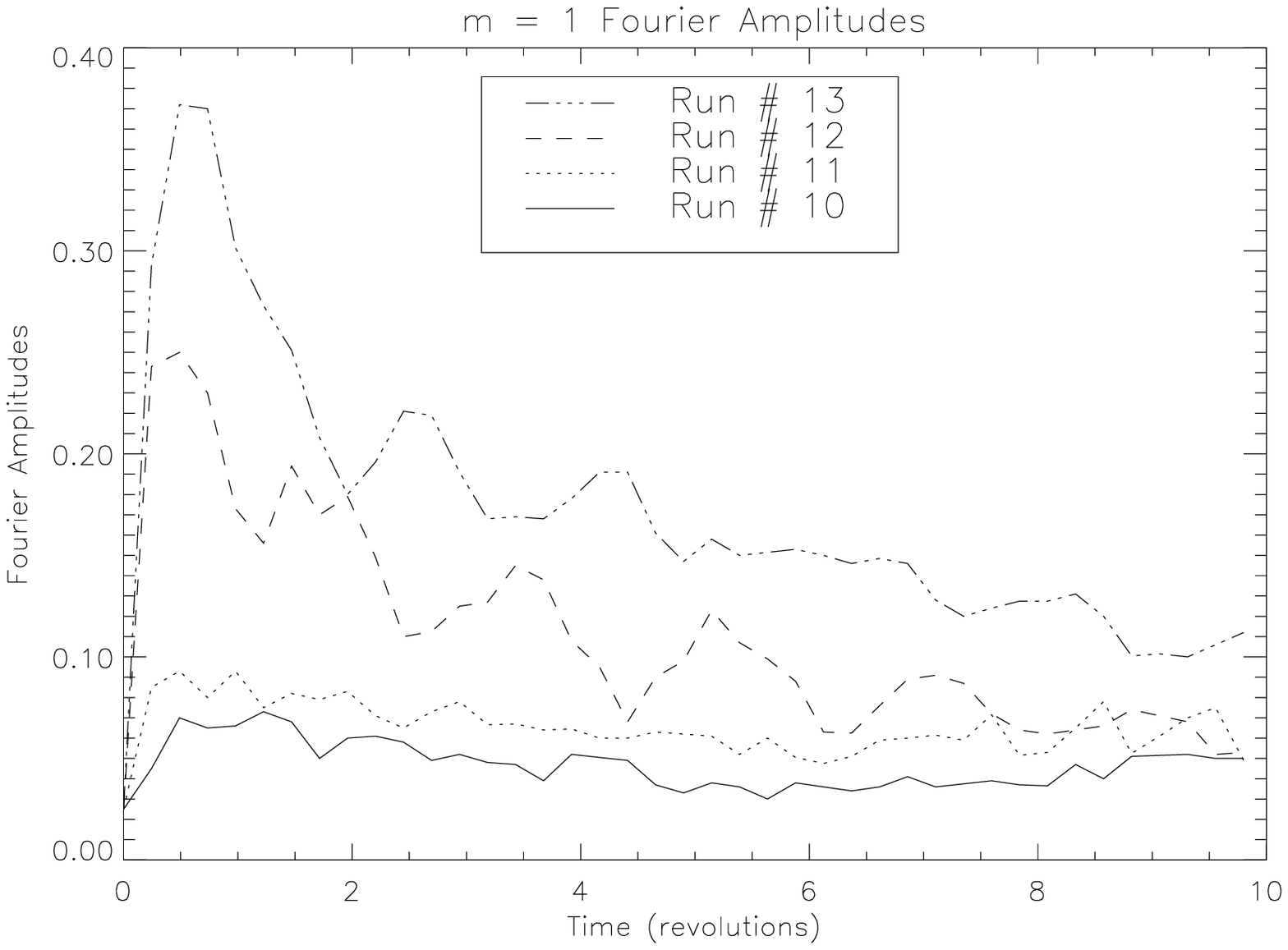]{The m=1 (one-armed)
behavior of 4 Sc runs labeled
10, 11, 12, and 13.\label{figure16}}

\figcaption[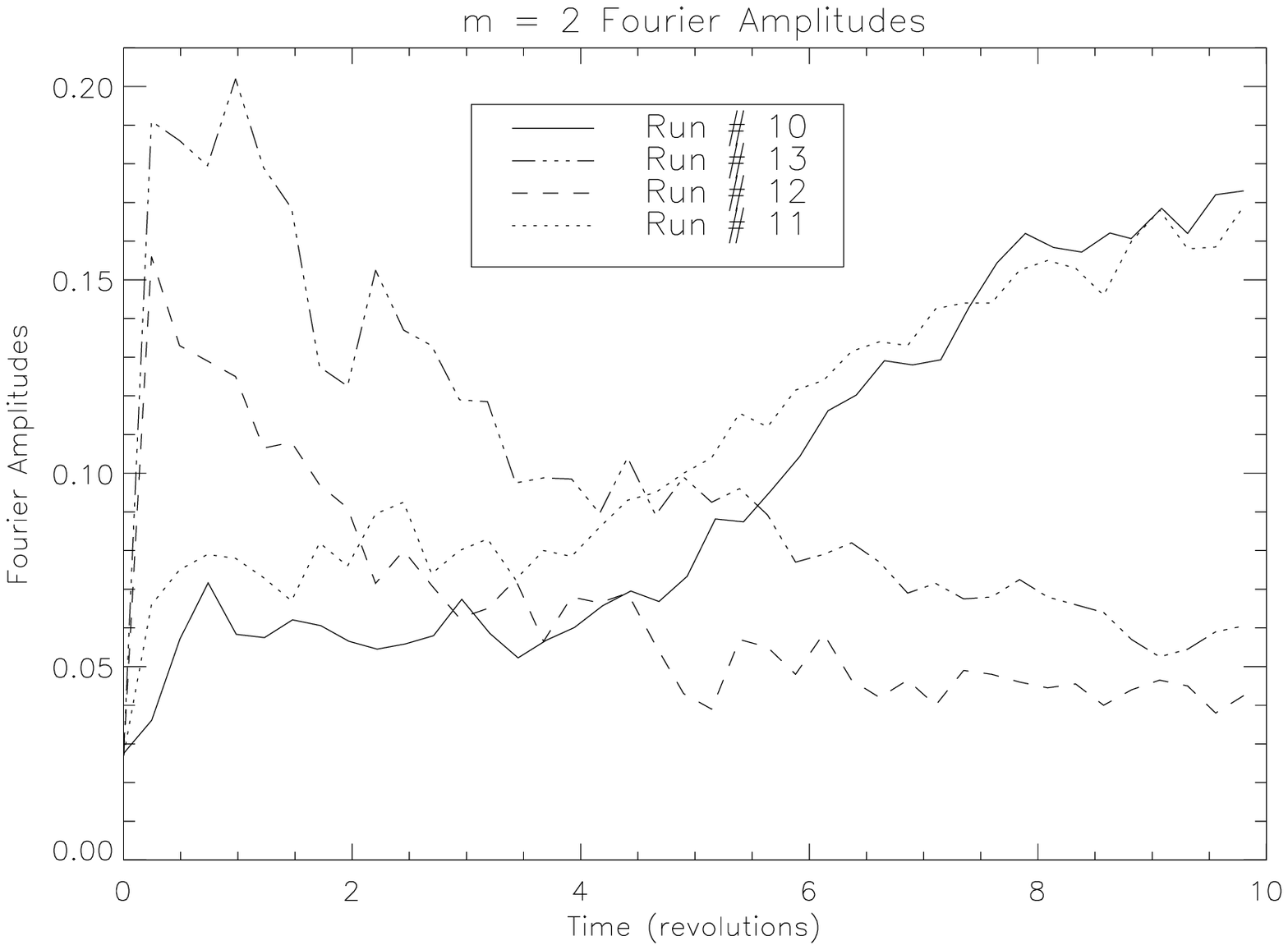]{The $m=2$ (bar) behavior of the same four
runs
used in Figure~\ref{figure16}.\label{figure17}}

\figcaption[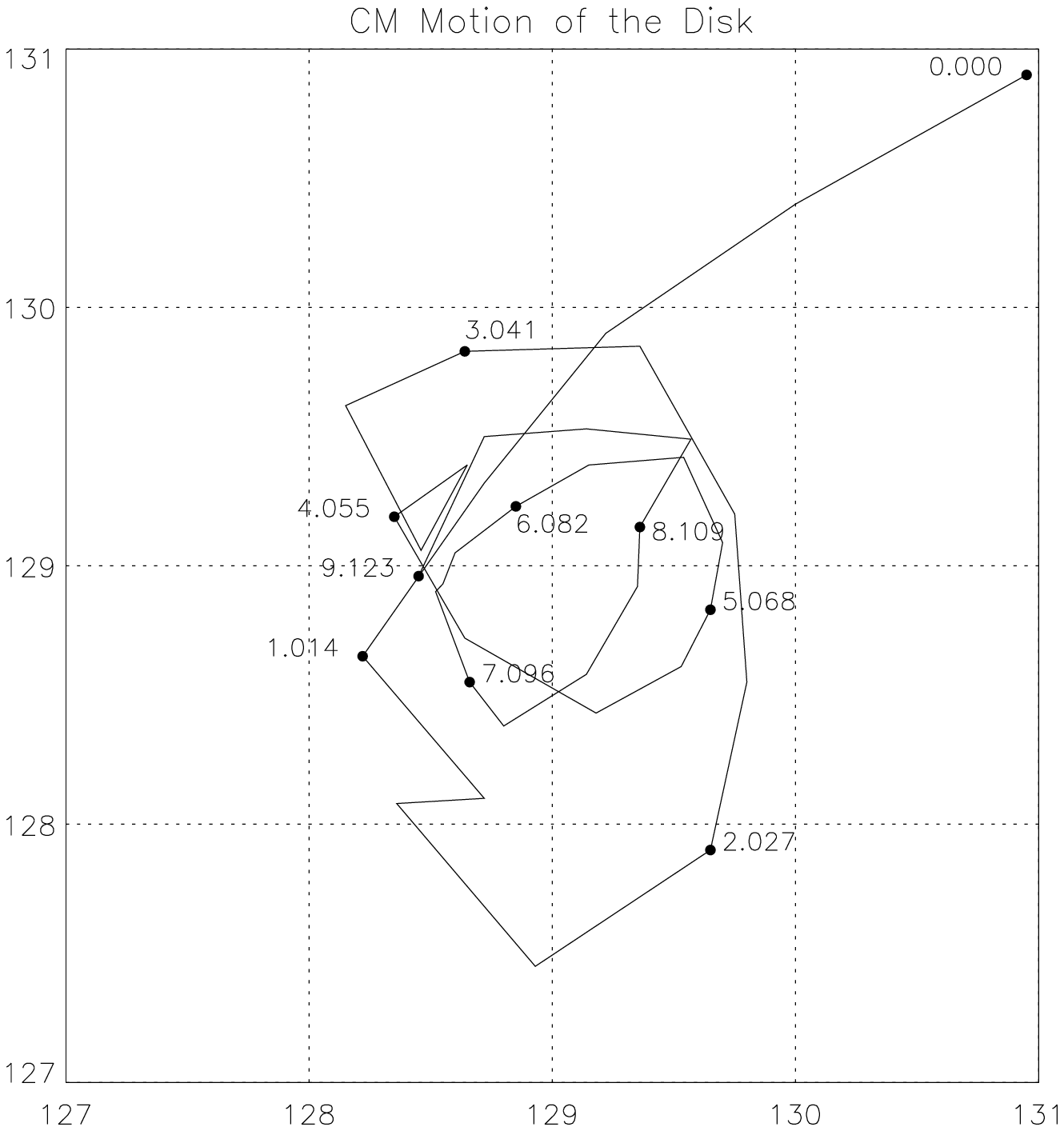]{Center of mass motion
of the disk in Run 11.
The halo in this run is fixed, with its center of mass
located at (129,129).
The numbers along the path are time units in revolutions.\label{figure18}}

\figcaption[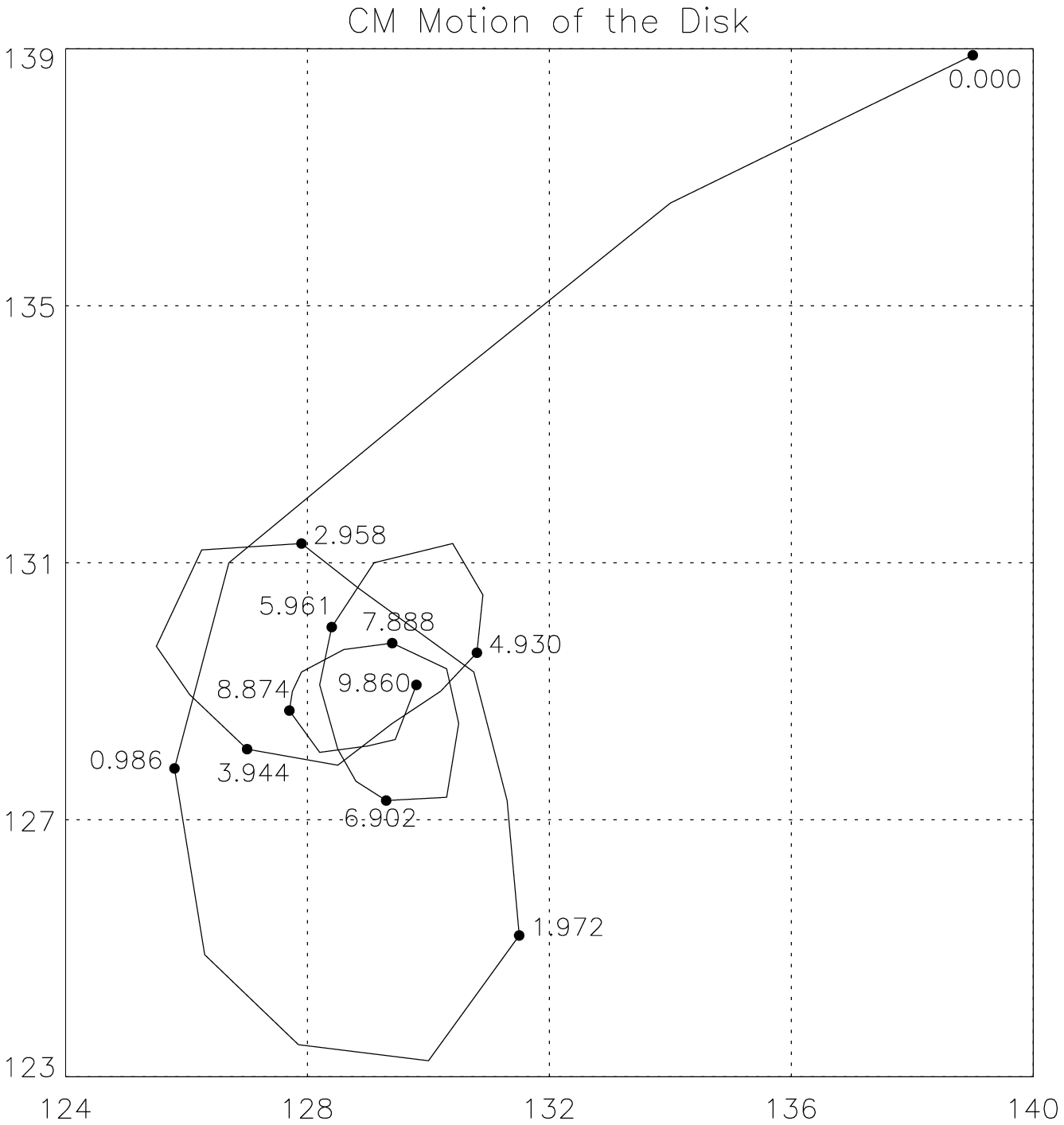]{Center of mass motion
of the disk in Run 12.
The halo in this run is fixed, with its center of mass
located at (129,129).
The numbers along the path are time units in revolutions.\label{figure19}}

\figcaption[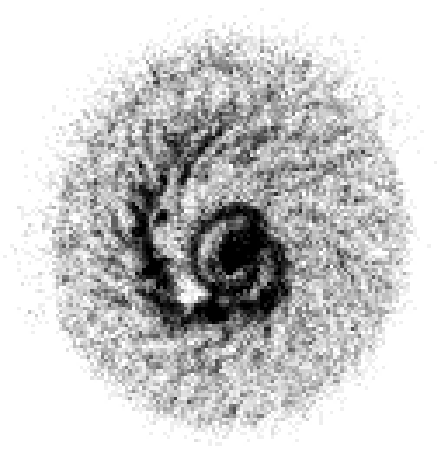]{Particle mass distribution
for Run 13 (Sc mass
distribution, large offset, high,
corotating angular momentum).\label{figure20}}

\figcaption[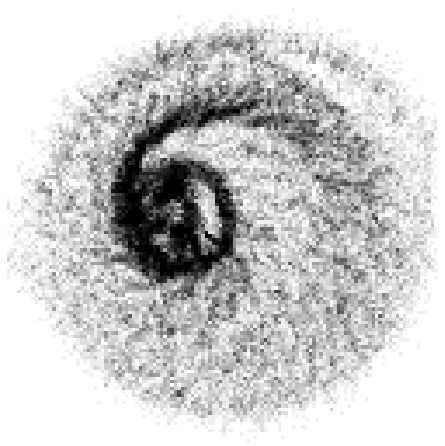,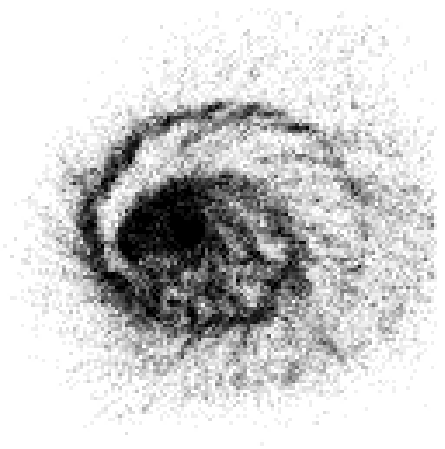,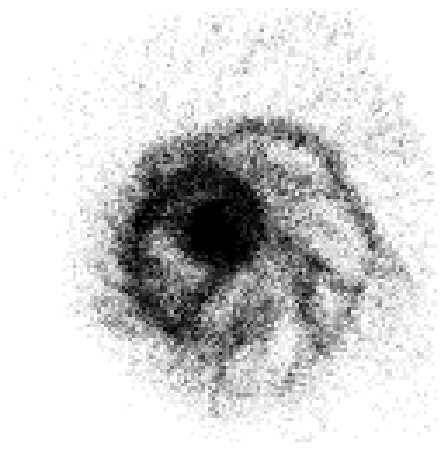]{Run 14 shows the initial
stationary trailing-arm spiral
expanding radially and then as a
leading arm spiral that persists
for the rest of the run.  (a) This
is at 0.42 rotation periods.  (b) At 0.83 rotation
periods.  (c) At 1.7 rotation
periods.
\label{figure21}}

\figcaption[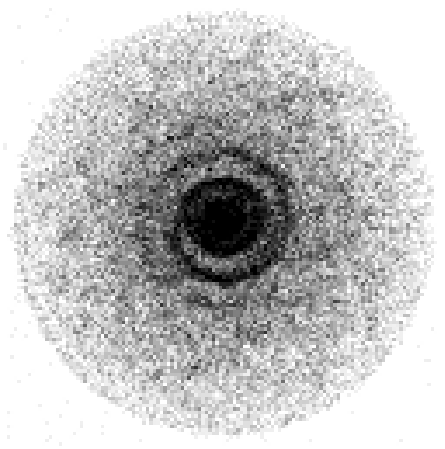,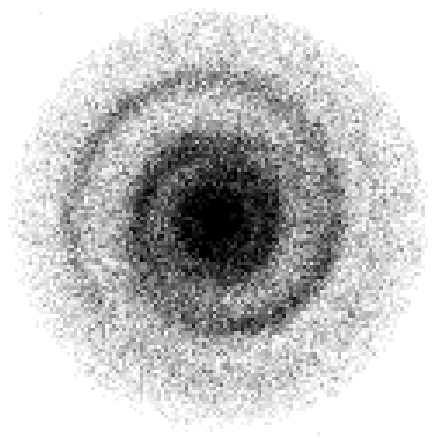,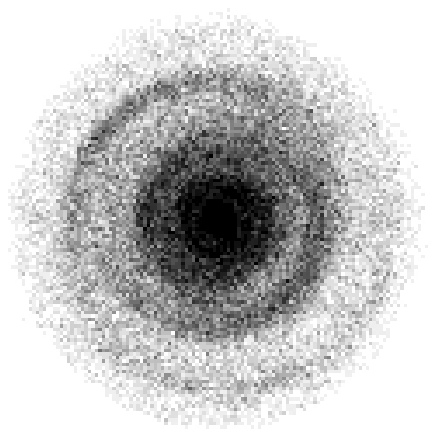]{Run 15 has equal numbers of
particles orbiting in opposite
directions and begins with a small
offset and low angular momentum.
(a) This is at $1.6$ rotation periods and shows an
initially tight one-armed spiral. (b) At
$5.0$ rotation periods: The spiral
expands.  (c) At $6.0$ rotation periods:
The spiral bifurcates into two
spirals.\label{figure22}}

\figcaption[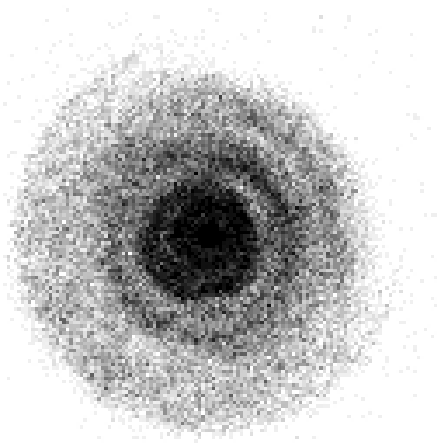]{This image from Run 16
(equal numbers of
particles orbiting in opposite
directions, a large initial offset
from the halo potential,
and low angular momentum) after $0.25$
rotation periods shows the
formation of a reflection-symmetric pattern.
This run also shows the formation
of non-axissymmetric structures.\label{figure23}}

\figcaption[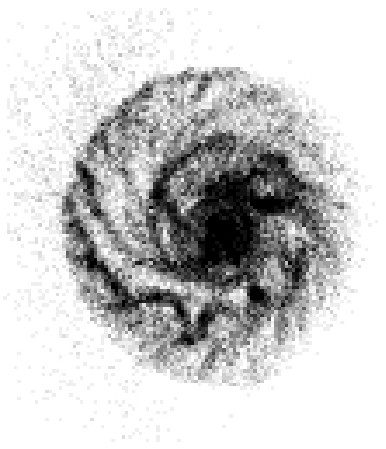]{Mass concentrations in
Run 20 (isothermal halo, black
hole, Sc disk, large offset,
high angular momentum).
These
transient features persist
for about 3 rotation periods.  Such
features are seen in spiral
galaxies such as M101.\label{figure24}}

\begin{deluxetable}{rllll}
\tabletypesize{\footnotesize}
\tablecolumns{5}
\tablewidth{0pc}
\tablecaption{Runs with Intruder, dynamic halo\label{tab1}}
\tablehead{
\colhead{Run \#} &
\colhead{Halo} &
\colhead{$Q_{0,halo}$} &
\colhead{$T_{0}$\tablenotemark{a}} &
\colhead{Description}}
\startdata
1\phm{1111} & $N$-body, dynamic & 0.3 &  & cold, clumpy \\
2\phm{1111} & $N$-body, dynamic & 5.0 &  & hot, clumpy \\
3\phm{1111} & Hydro, dynamic &  & $0.5\,$K & cold, smooth \\
4\phm{1111} & Hydro, dynamic &  & $5.0\,$K & cool, smooth \\
5\phm{1111} & Hydro, dynamic &  & $5\times10^{5}\,$K & hot, smooth \\
\enddata
\tablenotetext{a}{Measured at the half mass radius}
\end{deluxetable}

\begin{deluxetable}{rllll}
\tabletypesize{\footnotesize}
\tablecolumns{5}
\tablewidth{0pc}
\tablecaption{Dynamic Halo Simulations without Intruder\label{tab2}}
\tablehead{
\colhead{ } &
\colhead{Particle} &
\colhead{Halo} &
\colhead{Angular} &
\colhead{Final} \\
\colhead{Run \#} &
\colhead{Offset\tablenotemark{a}} &
\colhead{Type} &
\colhead{Momentum\tablenotemark{b}} &
\colhead{Bar?}}
\startdata
6\phm{1111} & $10\sqrt{2}$ & $N$-body & low, CCW & no \\
7\phm{1111} & $10\sqrt{2}$ & $N$-body & high, CCW & no \\
8\phm{1111} & $10\sqrt{2}$ & Hydro & low, CCW & yes \\
9\phm{1111} & $10\sqrt{2}$ & Hydro & high, CCW & no \\
\enddata
\tablenotetext{a}{This is the initial offset of the star particles from
the halo's center of mass.}
\tablenotetext{b}{The initial particle angular momentum can be either 0,
low, or high, and it can be either clockwise (CW) or counterclockwise
(CCW).}
\end{deluxetable}

\begin{deluxetable}{rlllll}
\tabletypesize{\footnotesize}
\tablecolumns{6}
\tablewidth{0pc}
\tablecaption{Simulations with a fixed halo\label{tab3}}
\tablehead{
\colhead{ } &
\colhead{Particle} &
\colhead{Angular} &
\colhead{Mass} &
\colhead{Counter-rotating} &
\colhead{Final} \\
\colhead{Run \#} &
\colhead{Offset\tablenotemark{a}} &
\colhead{Momentum\tablenotemark{b}} &
\colhead{Distribution\tablenotemark{c}} &
\colhead{Component?\tablenotemark{d}} &
\colhead{Bar?}}
\startdata
10\phm{1111} & 0 & 0 & Sc & \phm{11111}no & yes \\
11\phm{1111} & $2\sqrt{2}$ & low, CCW & Sc & \phm{11111}no & yes \\
12\phm{1111} & $10\sqrt{2}$ & low, CCW & Sc & \phm{11111}no & no \\
13\phm{1111} & $10\sqrt{2}$ & high, CCW & Sc & \phm{11111}no & no \\
14\phm{1111} & $10\sqrt{2}$ & high, CW & Sc & \phm{11111}no & no \\
15\phm{1111} & $2\sqrt{2}$ & low, CCW & Sc & \phm{11111}yes & no \\
16\phm{1111} & $10\sqrt{2}$ & low, CCW & Sc & \phm{11111}yes & yes \\
17\phm{1111} & 0 & 0 & Isothermal & \phm{11111}no & yes \\
18\phm{1111} & $10\sqrt{2}$ & low, CCW & Isothermal & \phm{11111}no &
yes\tablenotemark{e} \\
19\phm{1111} & 0 & 0 & Isothermal & \phm{11111}no & no \\*
   &   &   & plus black hole &  &  \\
20\phm{1111} & $10\sqrt{2}$ & high, CCW & Isothermal & \phm{11111}no &
yes\tablenotemark{f} \\*
   &              &           & plus black hole &  & \\
\enddata
\tablenotetext{a}{This is the initial offset of the star particles from
the halo's center of mass.}
\tablenotetext{b}{The initial particle angular momentum can be either 0,
low, or high, and it can be either clockwise (CW) or counterclockwise
(CCW).}
\tablenotetext{c}{This is the mass distribution of the static halo.
The mass distribution of the star particles was that of an Sc disk in
each run.}
\tablenotetext{d}{If counter-rotating components are present, then
50\% of the star particles rotate clockwise, and 50\% rotate
counterclockwise.}
\tablenotetext{e}{$m = 2$ (bar) Fourier amplitude 20\% lower 
than in Run 17.}
\tablenotetext{f}{Small bar only.}
\end{deluxetable}

\begin{figure} 
\figurenum{1} 
\plotone{f1.eps}
\caption{}
\end{figure}

\begin{figure} 
\figurenum{2} 
\plotone{f2.eps}
\caption{}
\end{figure}

\begin{figure} 
\figurenum{3} 
\plotone{f3.eps}
\caption{}
\end{figure}


\begin{figure} 
\figurenum{4} 
\plotone{f4.eps}
\caption{}
\end{figure}

\begin{figure} 
\figurenum{5} 
\plotone{f5.eps}
\caption{}
\end{figure}

\begin{figure} 
\figurenum{6} 
\plotone{f6.eps}
\caption{}
\end{figure}

\begin{figure} 
\figurenum{7} 
\plotone{f7.eps}
\caption{}
\end{figure}

\begin{figure} 
\figurenum{8} 
\plotone{f8.eps}
\caption{}
\end{figure}

\begin{figure} 
\figurenum{9} 
\plotone{f9.eps}
\caption{}
\end{figure}

\begin{figure} 
\figurenum{10} 
\plotone{f10.eps}
\caption{}
\end{figure}

\begin{figure} 
\figurenum{11a} 
\plotone{f11a.eps}
\caption{\label{figure11a}}
\end{figure}

\begin{figure} 
\figurenum{11b} 
\plotone{f11b.eps}
\caption{\label{figure11b}}
\end{figure}

\begin{figure} 
\figurenum{12a} 
\plotone{f12a.eps}
\caption{\label{figure12a}}
\end{figure}

\begin{figure} 
\figurenum{12b} 
\plotone{f12b.eps}
\caption{\label{figure12b}}
\end{figure}

\begin{figure} 
\figurenum{13a} 
\plotone{f13a.eps}
\caption{\label{figure13a}}
\end{figure}

\clearpage

\begin{figure} 
\figurenum{13b} 
\plotone{f13b.eps}
\caption{\label{figure13b}}
\end{figure}

\begin{figure} 
\figurenum{14} 
\plotone{f14.eps}
\caption{}
\end{figure}

\begin{figure} 
\figurenum{15} 
\plotone{f15.eps}
\caption{}
\end{figure}

\begin{figure} 
\figurenum{16} 
\plotone{f16.eps}
\caption{}
\end{figure}

\begin{figure} 
\figurenum{17} 
\plotone{f17.eps}
\caption{}
\end{figure}

\begin{figure} 
\figurenum{18} 
\plotone{f18.eps}
\caption{}
\end{figure}

\begin{figure} 
\figurenum{19} 
\plotone{f19.eps}
\caption{}
\end{figure}

\begin{figure} 
\figurenum{20} 
\plotone{f20.eps}
\caption{}
\end{figure}

\begin{figure} 
\figurenum{21a} 
\plotone{f21a.eps}
\caption{\label{figure21a}}
\end{figure}

\begin{figure} 
\figurenum{21b} 
\plotone{f21b.eps}
\caption{\label{figure21b}}
\end{figure}

\begin{figure} 
\figurenum{21c} 
\plotone{f21c.eps}
\caption{\label{figure21c}}
\end{figure}

\begin{figure} 
\figurenum{22a} 
\plotone{f22a.eps}
\caption{\label{figure22a}}
\end{figure}

\begin{figure} 
\figurenum{22b} 
\plotone{f22b.eps}
\caption{\label{figure22b}}
\end{figure}

\begin{figure} 
\figurenum{22c} 
\plotone{f22c.eps}
\caption{\label{figure22c}}
\end{figure}

\begin{figure} 
\figurenum{23} 
\plotone{f23.eps}
\caption{}
\end{figure}

\begin{figure} 
\figurenum{24} 
\plotone{f24.eps}
\caption{}
\end{figure}

\end{document}